\documentclass[superscriptaddress,twocolumn,10pt]{revtex4-1}
\usepackage[english]{babel}
\usepackage{amsmath,graphicx,color}
\usepackage{booktabs,array,dcolumn}

\usepackage[letterpaper,top=2cm,bottom=2cm,left=3cm,right=3cm,marginparwidth=1.75cm]{geometry}

\usepackage{soul}
\usepackage{caption}
\usepackage{subcaption}
\usepackage{siunitx}
\usepackage[colorlinks=true, allcolors=blue]{hyperref}

\newcommand{\lstar}{{L^{*}}}

\newcommand{\beginsupplement}{%
        \setcounter{section}{0}

        \setcounter{table}{0}
        \renewcommand{\thetable}{S\arabic{table}}%
        \setcounter{figure}{0}
        \renewcommand{\thefigure}{S\arabic{figure}}%
      }

\begin{document}

\author{Omar-Farouk Adesida} 
\affiliation{Department of Physics, University of Warwick, Coventry, CV4 7AL, UK}
\author{David Quigley}
\affiliation{Department of Physics, University of Warwick, Coventry, CV4 7AL, UK}
\author{ Livia B. P\'artay}
\affiliation{Department of Chemistry, University of Warwick, Coventry, CV4 7AL, UK}

\title{Revisiting the Phase Diagram of Hard Sphere Dumbbells with Nested Sampling: Known Phases and New Packing Variants}


\date{\today}

\begin{abstract}

We explore the use of the nested sampling technique to sample the configuration space of non-spherical hard particles. We employ the technique on the hard dumbbell system consisting of two hard spheres connected by a rigid bond, and investigate the phase stability across a wide pressure range and for bond lengths from completely overlapping to tangential hard spheres. Nested sampling recovers all previously identified features of the phase diagram and identifies a family of new packing variants. The fluid phase, plastic crystal, close packed solid phases and aperiodic crystal are all sampled, and the transition points between these are mapped. Our results show good agreement with predictions made by existing equations of state, and former Monte Carlo simulations. Nested sampling also identified a close packed structure with $Pnma$ symmetry which has not previously been considered.

\end{abstract}

\maketitle

\section{Introduction}



Hard sphere-based models offer a convenient and simple way to capture a range of properties in both atomistic and molecular systems. While hard sphere-based models are computationally cheap to evaluate they are also capable of describing complex behaviour, such as the fluid-solid transition\cite{davidchack_direct_2000,noya2008determination,royall2023colloidal} (observed in some of the first atomistic simulations of hard-disks and hard-spheres \cite{alder_phase_1962, alder_phase_1957}), solid polymorphism \cite{radu_solid-solid_2009} and amorphous phases\cite{pusey_hard_2009,charbonneau_memory_2021}.
Systems of hard particles have also been used to develop a better understanding of the jamming phenomena\cite{ohern_jamming_2003}, glass transition and liquid crystal phases in materials \cite{charbonneau_memory_2021,whittle_liquid_1991}. 
With anisotropy introduced (for example via constructing unions of hard spheres), hard particle models have also been regularly used to represent colloidal systems, and demonstrate a wide variety of properties which depend on the shape and relative orientation of particles.\cite{tasios_glassy_2014,avendano2012phase,avendano2017packing,miller2010phase,anderson2020hoomd} 

Hard sphere models are also a popular system for validating and benchmarking simulation methods and configurational sampling algorithms in the context of soft matter. One such approach is nested sampling (NS) \cite{skilling_nested_2006,ashton_nested_2022}. This algorithm performs an unbiased and exhaustive exploration of the configuration space, allowing predictions of the thermodynamically relevant phases and evaluation of the partition function\cite{partay_efficient_2010,partay_nested_2021}. 
Nested sampling has been successfully applied to a range of materials, often discovering previously unconsidered phases or transitions, giving unique insight into features of the potential energy landscape as well as the behaviour of widely employed model potentials\cite{partay_nested_2014,baldock_constant-pressure_2017,dorrell_pressuretemperature_2020,gola_embedded_2018,bartok_insight_2021}. The NS approach is particularly well suited to high-throughput generation of phase diagrams as a function of model parameters, requiring no prior knowledge of the expected phases. However, previous applications of NS have focused almost exclusively on the study of spherically symmetric potentials and atomic models. 

\begin{figure}
     \centering
         \includegraphics[width=0.45\textwidth]{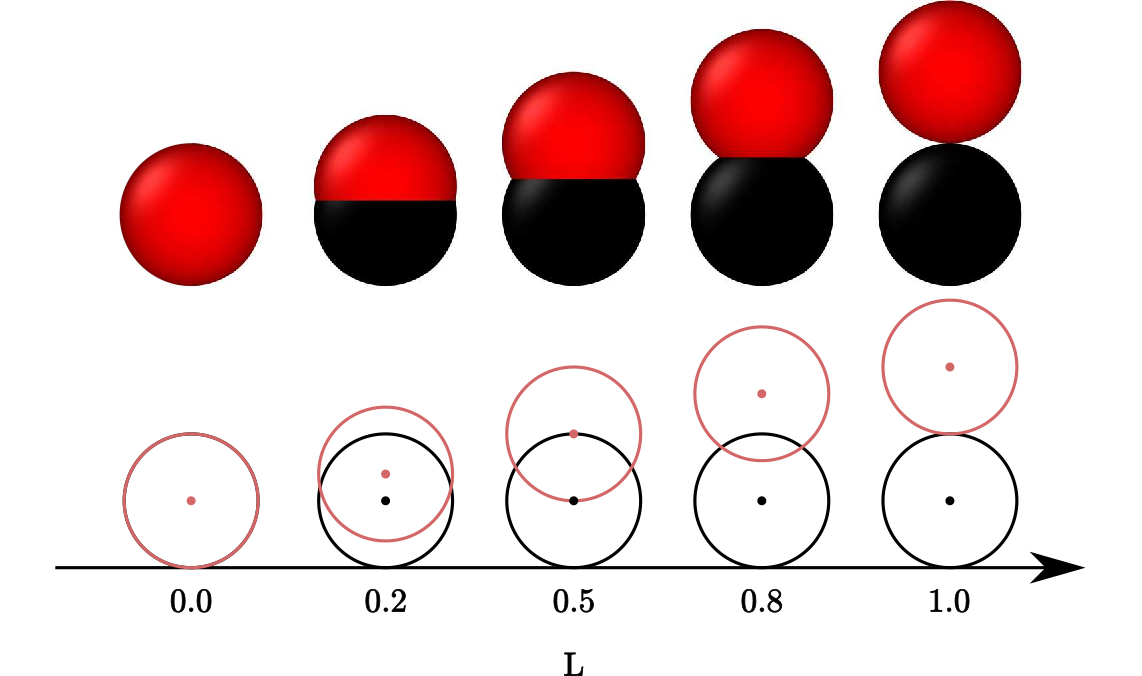}
         \caption{Shape of dumbbell particles with placing the hard spheres at different distances, $L$, normalised by the diameter of the constituting spheres. $L=0.0$ represents the simple hard-sphere system, while in case of $L=1.0$ the two spheres has only a single point of contact. Colouring is just for clarity.}
         \label{fig:dumbbell_example}
\end{figure}

In this paper we explore the application of NS to the hard-sphere dumbbell model in three dimensions, demonstrating the generalisation of the method to rigid molecular models, and hence to systems with rotational degrees of freedom. The dumbbells are constructed as the union of two hard spheres of radius $\sigma$ (we set $\sigma=1$ for all results in this study) connected by a rigid bond of length $L$. We vary $L$ to create a series of models from completely overlapping ($L=0$, a single hard sphere) to tangential spheres ($L=1$) as demonstrated in Figure~\ref{fig:dumbbell_example}. 

These particles have been studied for close to 50 years as an important model for small non-spherical molecules and colloids. They provide crucial insight into how thermodynamic properties depend on the geometry of their constituent particles, without the complication of energetic considerations, such as balance between short and long range interactions. As athermal models, the thermodynamically stable phase is determined entirely by the dimensionless pressure $P^{*} = P\sigma^3/k_{B}T$, where $P$ is the external pressure and $k_{B}$ is Boltzmann's constant. Phase transition pressures at any temperature $T$ are collapsed to the same dimensionless pressure by this scaling. Since we set $\sigma=1$ the dimensionless pressure is $P^{*}=P/k_{B}T$. 

Existing studies, primarily employing Monte Carlo \cite{vega_solidfluid_1992} and Molecular Dynamics techniques \cite{ni_crystal_2011}, have established a consensus on the main features of the hard dumbbell phase diagram in the $P^{*}-L$ plane. At $L=0$ the hard sphere phase diagram is recovered in which the fluid solidifies into a crystal of close packed spheres at approximately $P^{*}=11.54$ \cite{noya2008determination} via a first order transition. The face-centred-cubic (fcc) polytype is marginally more stable than the hexagonal-close-packed (hcp) \cite{PhysRevE.61.906}, however, pure fcc structure would only be expected if compression is sufficiently slow that hcp stacking faults are annealed away during crystallisation. When  $L$  is small but has non-zero value the solid phase becomes a plastic crystal (PC) first observed via direct simulation of crystallisation from the fluid \cite{Freasier01011976}. In the PC phase bond centres occupy the sites of a close packed lattice \cite{singer_monte_1990,vega_plastic_1997}. The dumbbells rotate around these bond centres but with clear deviations in their orientational correlation function from that of an ideal free rotor \cite{allen_molecular_1987}. Highly accurate free energy calculations based on a combination of lattice switching Monte Carlo and thermodynamic integration have been able to show that the most stable arrangement of dumbbell bond centres transitions  on increasing $L$ from fcc to hcp at all relevant densities before $L\leq0.15$ \cite{marechal_stability_2008}, leading to an analogy between the hcp PC and the $\beta$-N$_{2}$ phase of solid nitrogen.

Upon increasing $L$ further, \citeauthor{singer_monte_1990} \cite{singer_monte_1990} noted that the PC--fluid coexistence curve terminates at approximately $L=0.4$, suggesting that some other solid phase becomes more stable past this limit. Several candidates for this structure were first proposed by \citeauthor{vega_solidfluid_1992} \cite{vega_solidfluid_1992} including the $\alpha$-N$_2$ structure of solid molecular nitrogen (the orientationally ordered version of the fcc PC phase) and three other close packed structures. The latter were constructed as stacked bilayers in which both spheres comprising the dumbbell are arranged into close packed two-dimensional layers. The bonds that connect the two spheres in each dumbbell between these layers are tilted at an angle of $\arcsin{(L/\sigma\sqrt{3}})$ to the direction normal to these layers which ensures close packing within the bilayer. Subsequent bilayers can be stacked onto the first with a $\sigma/\sqrt{3}$ shift that is either parallel or antiparallel to the in-layer component of the dumbbell bond vectors. The former leads to a structure which is equivalent to fcc (ABC)$_n$ stacking of hard spheres when $L=1$ and the latter leads to hcp (AB)$_n$ stacking when $L=1$. \citeauthor{vega_solidfluid_1992} named these two cases CP1 and CP2 respectively. A third structure in which the direction of the dumbbell bond vectors alternates between bilayers was named CP3. This corresponds to (ABCBC)$_n$ stacking at $L=1$. Other stacking sequences of these dumbbell layers are of course possible (see supplementary information section \ref{sec:CPx}), including a dhcp polytype i.e. (ABCB)$_n$, random stacking disordered sequences, and structures in which the bond vector connecting the two spheres in the dumbbell rotates around the normal to the stacked layers in steps of $60^{\circ}$, exploiting the six-fold symmetry of the layers. 
We refer to all these stacked layers structures collectively as CPx. 

\citeauthor{vega_stability_1992} also showed that the CP1 structure becomes more stable than the PC structure at approximately $L=0.38$, and for any smaller $L$ at higher pressures. This has been confirmed by subsequent studies \cite{marechal_stability_2008,sweatman_self-referential_2009} as has their observation that the CP1, CP2 and CP3 polytypes have near-identical stability \cite{kowalik_free_2008}. The $\alpha$-N$_2$ structure (analogous to the fcc PC crystal but with bonds aligned along the diagonal of the cubic cell) was found to never be thermodynamically stable relative to the higher density close packed phases, but is mechanically stable for small $L$ and does not disorder into the PC phase when simulated at high pressures. 

Using free energy calculations based on Monte Carlo simulation, \citeauthor{vega_stability_1992} \cite{vega_stability_1992} were also able to compute the pressure at which CP1, CP2 and CP3 coexist with the fluid for three values of $L$. We note here that studies of crystal nucleation in the hard-sphere dumbbell system have either avoided the regime of $L$ for which the CPx structure putatively represents the stable solid phase or been unable to capture its formation due to a combination of high-energy barriers and slow dynamics \cite{ni_crystal_2011,zubieta_nucleation_2019}. This potentially presents a significant challenge to the NS algorithm which relies on identification of solid phases by evolution of samples starting in the melt. To our knowledge no simulation study has directly captured crystallisation from the fluid into these close packed structures. We are also unaware of any simulations which capture phase coexistence between the fluid and any of the CPx structures. 

As $L$ approaches 1.0 a further phase transition has been shown to occur. As noted above, in this limit the CPx phases reduce to stacking of close packed spheres and there is no longer any need for the dumbbell orientations to align to maintain that packing. By analogy with the 2D aperiodic phase of \citeauthor{wojciechowski_nonperiodic_1991} \cite{wojciechowski_nonperiodic_1991}, \citeauthor{vega_solidfluid_1992} \cite{vega_solidfluid_1992} considered a phase in which the dumbbell spheres are close packed (specifically as per their CP3 structure) with a random network of bonds. The additional entropy of the bond network \cite{nagle_new_1966} in this aperiodic phase (AP) enhances its stability relative to the orientationally ordered CPx phases. The presence of a phase transition into the AP phase has been confirmed by multiple free energy calculations \cite{kowalik_free_2008,marechal_stability_2008}. This phase also presents a potential challenge to NS. Many realisations of the disordered bond network will need to be located and sampled by the algorithm in order to accurately locate its phase boundaries.

In the present work we apply NS to this system and demonstrate that it can accurately identify and sample the full range range of complex phases without prior knowledge of their structures, and provide insight into thermodynamic properties across phase transitions. Furthermore, it identifies a family of previously unconsidered close-packed dumbbell structures for intermediate $L$. Our work also highlights challenges specific to hard potentials, such as jamming, and the complexity of the configuration space of particles with internal degrees of freedom. These results serve as important first step towards using NS to explore the phase behaviour of molecular systems in the future.


\section{Methodology}

\subsection{Hard Dumbbell Model}

In the hard dumbbell model, particles are constructed using two spheres, connected by a rigid bond of fixed length $L$. Two spheres within the same particle do not interact with each other. All other pairs interact through the hard-sphere potential, $U(r)$, in which the potential energy is either zero or infinity, thus  
\begin{center}
$$ 
U\left( r \right) = \left\{ \begin{array}{lll}
\infty & ; & r <  \sigma \\
0      & ; & r \ge \sigma \end{array} \right.,
$$ 
\end{center}
where $\sigma$ is the diameter of a sphere within the dumbbell, and $r$ is the distance between the centre of two spheres (not within the same particle).

Our nested sampling implementation implements this potential using a modified version of the \texttt{hs\_alkane} package \cite{quigley_hs_alkane_2019}, which had been used in a previous work by Quigley and Bridgwater~\cite{bridgwater_lattice-switching_2014}. 

\subsection{Monte Carlo moves}


We define  a Monte Carlo (MC) sweep as the number of MC moves equal to the degrees of freedom in the system. We attempt five MC move types. The first changes the volume of the simulation cell box isotropically. In order to be able to accommodate different crystalline symmetries in the simulation cell, not only volume steps, but isometric stretch and shear steps were also introduced to be able to change the aspect ratio and tilt of the box. These two types of move change the aspect ratio of the box while maintaining the cell angles or change the cell angles while maintaining the cell lengths. We also include dumbbell translation and rotation moves. These five moves are attempted with random probability in the ratio $1:3:3:3N:2N$  respectively.

Translation moves perturb the centre-of-mass position of a particle by a random vector of magnitude $0$ to $\Delta r_{\mathrm{max}}$.  For rotation moves we perturb the orientation of a particle by choosing a random axis through the centre-of-mass of the particle and then rotating it about that axis by $\alpha$ angle chosen randomly in the range between $-\alpha_\textrm{max}$ and $+\alpha_\textrm{max}$.
Each rotation or translation move is attempted on a randomly selected particle and accepted if no overlaps between hard spheres are generated. Moves which perturb the shape of the simulation box are rejected if overlaps are created and otherwise accepted.
Moves that change the volume are rejected if overlaps are generated, otherwise accepted with probability
\begin{multline}
\mathcal{P}_{\mathrm{acc}}(V_o\rightarrow V_n) = \\ = \exp{\left[-\beta P(V_{n}-V_{o}) + N\log{(V_{n}/V_{o})}\right]}
\end{multline}
where $V_o$ and $V_n$ are the old and new volumes, respectively, and $\beta$ is the inverse temperature and $P$ is the pressure applied to the system of dimers. We set the inverse temperature $\beta=1$ in all simulations. 

The size of the proposed perturbations are chosen such that the acceptance probability of a move remains between 20 and 50\%. Since the stepsizes strongly depend on the density of the phase, these were adjusted during the equilibration of the Monte Carlo simulations, as well as periodically updated during nested sampling. In the latter case these adjustments are performed every $K/2$-th nested sampling iterations, where $K$ is the number of walkers (see below). Theses adjustments are performed during a short simulation that does not contribute to the nested sampling statistics.

\subsection{Nested Sampling}


Nested sampling (NS) is an iterative ``top-down'' approach, starting the sampling at the high-energy region of the phase space and progressing towards the ground state structure. This is done via using a large set of independent configurations, and contracting the available phase space volume with a fixed ratio during the iterative steps~\cite{partay_nested_2021}.

Contrary to most of previous materials applications of the method, in case of the hard-sphere dumbbell particles the energy of the system is either zero or infinity, meaning that it is not a suitable choice to drive the exploration. Instead, we use the density of the configurations~\cite{partay_nested_2014}, thus progressing towards smaller volumes during the sampling, rather than lower energies. 
In the following we describe details of the nested sampling method as employed in the present work. 

The sampling starts by randomly generating $K$ configurations of the $N$ dumbbell system, each generated by placing $N$ number of dumbbells in the simulation box at random positions and random orientations, but ensuring no particles overlap. These initial configurations have a sufficiently large volume, such that they correspond to a low density fluid phase in the ideal gas limit.
These $K$ configurations form the so-called \emph{live set} with each configuration constituting a \emph{walker}.  During the sampling, at each iteration $i$, the walker with the largest volume is recorded, $V_i = \max\{V_1 ... V_K\}$, and replaced with a new walker generated by performing a series of the Monte Carlo sweeps, described above, as well as the shape and size of the simulation cell, on a randomly picked existing walker. 
During this Monte Carlo walk every step is accepted, unless the step generates an overlap or causes the volume to exceed that of the old configuration, i.e. the $V_\mathrm{new} < V_i$ criteria has to be always satisfied. Due to the small size of the system typically used in NS, we also have to introduce an additional criteria to prevent very anisotropic cells to be favoured in early stages of the sampling. Such cells limit position fluctuations along the short direction and affect the sampling and formation of the crystalline phases by introducing unphysical correlations \cite{Frenkel_cell,baldock_determining_2016}. Hence we restrict the shortest cell height to remain larger than 80\% of a cube of the same volume. This process is repeated iteratively until all the $K$ configurations have reached a volume low enough (i.e. high enough packing fraction), such that we have generated samples representative of the the ground state structure or structures. 
Using the volumes of the culled configurations, the partition function, $\mathcal{Z}$, can be calculated at a given pressure $P$, as
\begin{equation}\label{eq:Z}
\mathcal{Z}(P) = \sum_i w_i e^{-\beta PV_i} , \\
\end{equation}
where $\beta = 1/(k_BT)$ is the thermodynamic inverse temperature, and $w_i$ is the NS weight,
\begin{equation}\label{eq: weight}
    w_i = \Gamma_{i-1} - \Gamma_i = \left(\frac{K}{K+1}\right)^{i-1} -\left(\frac{K}{K+1}\right)^i,
\end{equation}
where $K$ is the number of walkers and $\Gamma_i$ is the phase space volume fraction at the $i$th iteration, with $\Gamma_0=1$. 
The isothermal compressibility, $\kappa$, can be derived from the partition function, using the relations discussed in Ref.~\cite{grigoriev_monte-carlo_2020}, at an arbitrary pressure, once the sampling is finished,
\begin{multline}
       \kappa = \frac{\left<\delta V^2\right>}{\left< V \right>k_BT} = -\frac{1}{V}\Big(\frac{dp}{dV}\Big)_T = \\
    =\frac{\mathcal{Z}\sum_i V_i^2 w_i e^{-\beta PV_i} - \left(\sum_i V_i w_i e^{-\beta PV_i}\right)^2 }{k_BT \mathcal{Z} {\sum_i V_i w_i e^{-\beta PV_i}}}. 
\end{multline}
Similarly, the expected value of any observable can be calculated from the appropriately weighted average of the samples generated during the NS. For example, the expected value of volume can be calculated by
\begin{equation}\label{eq:volume-pressure}
    \left<V\right>(P) = \frac{\sum_i V_i w_i e^{-\beta PV_i}}{\mathcal{Z}(P)}. 
\end{equation}

Thus, a single NS run can be performed to explore the configuration space over a continuous pressure range. Peaks in the compressibility indicate the location of phase transitions, i.e. pressures at which the volume of the system changes very rapidly. 
Varying the separation distance between the two spheres composing the dumbbell, $L$, in different NS calculations, we are able to generate the pressure--bond-lengths phase diagram of hard-sphere dumbbells.

    



\subsection{Computational details}

\subsubsection{NS parameters}
Initial configurations for the NS live set were generated within a cubic cell of volume $N \times 30\sigma^3$ for each walker, where $N$ is the number of dumbbells. 
We used either $N=16$ or $N=32$ in the present work. While this means we use very small system sizes, previous works have shown that this does not significantly affect the sampled phases, and with e.g. the solid-liquid transition usually underestimated only by 3-10\%~\cite{baldock_determining_2016}. It is also worth noting that, thanks to having multiple configurations sampling the phase space simultaneously, NS does not require that phase transitions are sampled with both phases present in the same simulation cell, and hence no interfacial configurations have to be accommodated.
The number of walkers, $K$, controls the resolution of the sampling, with the computational cost depending on it linearly. We chose $K$ such that the resulting compressibility peaks are sufficiently converged (i.e. the deviation of the peak location in three independent runs varied by less than 5\%, for examples see Figures~\ref{fig:compressibility_02} and~\ref{fig:comp2}). We found that $K=1152$ is sufficient for the majority of the studied systems to achieve this. This resolution typically required $10^6$ NS iterations to reach configurations representative up to $150 \sigma^3/k_BT$ pressure after starting the sampling from the random initial configurations.  

During the NS iteration, new configurations were generated by cloning a randomly selected existing walker and using 4800 MC sweeps to decorrelate it. 
Simulations were parallelised as described in Ref.~\cite{partay_nested_2021}, spreading the cost of this decorrelation across multiple iterations. During a parallel nested sampling run with using $N_\mathrm{proc}$ number of processors, not only the freshly cloned configuration is decorrelated by an MC walk, but the $N_\mathrm{proc}-1$ extra processor are used to each propagate one other randomly selected walker. These MC walks are shorter, in our case $4800 \mathrm{~sweeps}/N_\mathrm{proc}$, as every configuration is propagated during multiple NS iterations, each achieving the required decorrelation length on average.

The nested sampling code used in the current work is available on the github repository provided in \cite{adesida_hans_2023}.
This also requires the use of the $\mathtt{hs\_alkane}$ package developed by Quigley and Bridgwater~\cite{quigley_hs_alkane_2019}, an implementation of the hard-sphere alkane model of Malanoski and Monson~\cite{malanoski_solid-fluid_1999} which reduces to a dumbbell model for a chain length of two. 

\subsubsection{Additional MC simulations}

A series of traditional Metropolis Monte Carlo (MC) simulations in the $NPT$ ensemble were also performed at a range of different pressures in order to explore select configurations produced by nested sampling under equilibrium conditions. Specifically, these were used to explore order parameter distributions and autocorrelation functions in crystalline solid structures identified by nested sampling. These reference simulations were performed using 256 or 432 dumbbells and performing \num{5e5} MC sweeps, consisting of a combination of translational, rotational and cell moves (affecting both the volume and shape of the simulation box). The first \num{1e6} MC sweeps were always used to equilibrate the system. Configurations were output every 100 MC sweeps afterwards, such that a total of \num{5e3} configurations were available for analysis.

In addition, longer simulations were performed to confirm that other structures found during nested sampling had equivalent densities to those already established in the literature (see supplementary information section \ref{sec:CPpf}). These simulations used 960 dumbbells, with an equilibration period of \num{5e5} MC sweeps, followed by sampling at intervals of 20 sweeps for an additional \num{3e6} sweeps. 


\section{Results}

\begin{figure}[h]
    \centering
            \includegraphics[width=\linewidth]{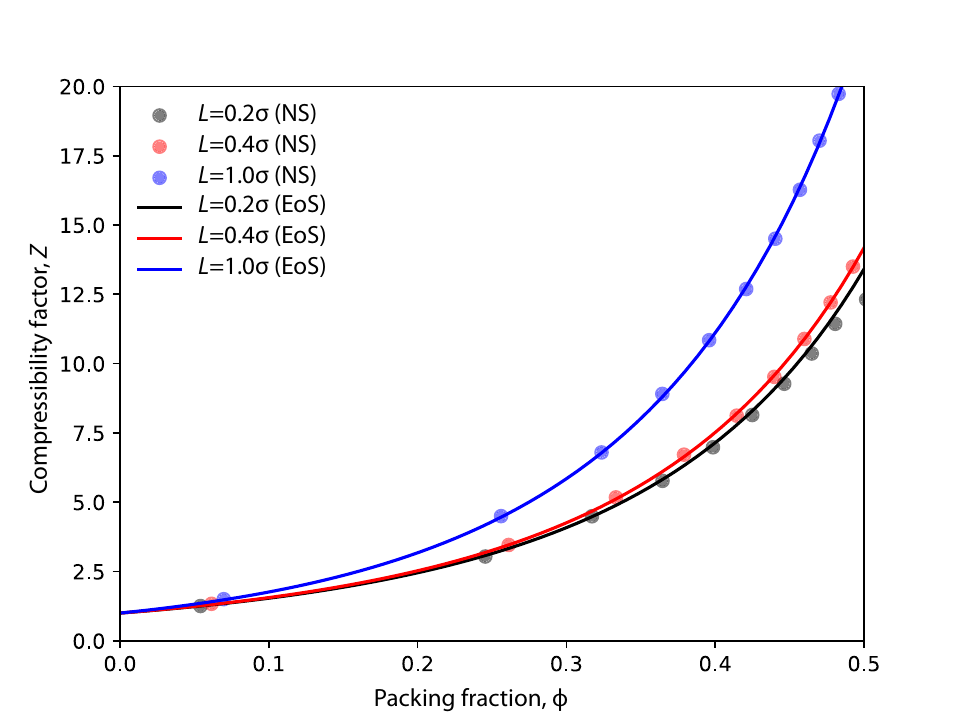}
    \caption{Compressibility factor as a function of packing fraction, calculated by the equation of state (solid lines) from Ref.~\cite{vega_linear_1994} and from nested sampling (solid circles), for three different bond lengths in the fluid phase.}
    \label{fig:EoS}
\end{figure}

While there is no exact theoretical solution for the equation of state (EoS) of the hard dumbbell system, previous literature has approximated the equation of state allowing the calculation of the compressibility factor in the fluid phase, $Z=Nk_BT/(PV)$, as a function of packing fraction, $\phi$, for different bond length values~\cite{vega_linear_1994,boublik_equation_1977,Tildesley01091980}. 
In the current work we define the packing fraction as the total volume of the dumbbells divided by the volume of the simulation box. Note that some works use a reduced packing fraction, normalised by the radius of a sphere having the same volume as one dumbbell. 
In Figure~\ref{fig:EoS} we compare the compressibility factor of the fluid phase, calculated from our nested sampling simulations (volumes calculated by eq.~\ref{eq:volume-pressure})  to those calculated by the EoS suggested by Vega et al., showing excellent agreement.

Isothermal compressibilities, $\kappa$, were obtained from the partition function, using NS for different values of bond lengths in the range of $0.0 < L < 1.0$. Peaks on the compressibility curve, as well as changes in the gradient of the density were used to locate phase transitions in the systems.
In the following three sections we present details of the observed phases and transitions, with different bond length ranges showing similar behaviour discussed together. 

\subsubsection{Bond lengths $0.0\sigma < L \leq 0.4\sigma$} 

\begin{figure}[h]
    \centering
            \includegraphics[width=\linewidth]{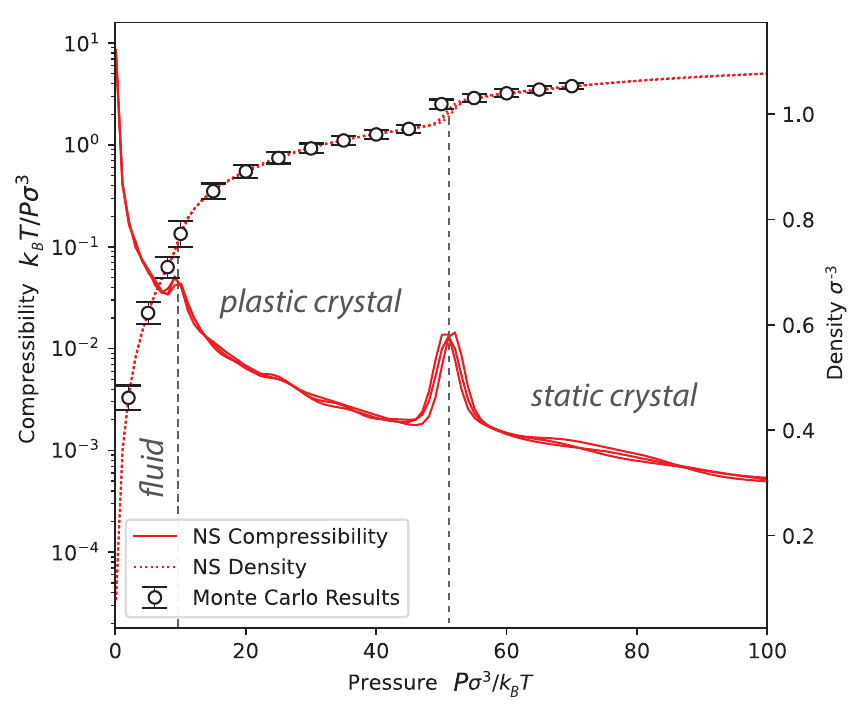}
    \caption{Compressibility (solid lines) and density curves (dashed lines) plotted for three parallel nested sampling runs at bond length $L=0.2$, as a function of pressure. The peaks in the compressibility curves, and the changes in gradient for the density observed at the same pressure values signal phase transitions in the system (marked by vertical grey dashed lines). Included are results of a MC simulation of the system under a range of relevant pressures, shown by open circles.}
    \label{fig:compressibility_02}
\end{figure}

For short bond lengths of $0 < L \leq 0.3$, two peaks were observed on the compressibility curves, indicating that there are three thermodynamically stable phases of these systems. This is contrast to the work of \citeauthor{vega_stability_1992} \cite{vega_stability_1992} in which only the plastic crystal and the fluid were identified as thermodynamically stable. We attribute this to their work only considering lower pressures. 
This is demonstrated in \autoref{fig:compressibility_02} in case of dumbbells of bond length $L=0.2$, showing the results of three independent parallel NS runs, with the alignment of the peaks also demonstrating the level of convergence of our sampling. For reference \citeauthor{vega_stability_1992} only considered pressures up to $Pd^3/k_{B}T=50$ where $d$ is the diameter of sphere that possesses identical volume to that excluded by the dumbbell. In our units and for $L=0.2$ this corresponds to  $P\sigma^3/k_{B}T=38.6$, significantly lower than the transition to a static phase.
In~\autoref{fig:compressibility_02}, we also show the expected value of density as a function of pressure, both calculated by nested sampling and reference MC simulations, showing very good agreement. 

In order to characterise the emerging solid phases, we use two order parameters to describe the translational and orientational order of the dumbbells. 
As translational order parameter we use the Steinhardt bond order parameter~\cite{QW}, $Q_6$, calculated over the centres of mass of the dumbbells, averaging the local order of all particles within a configuration. Taking into account the first coordination shell, the value of $Q_6$ increases slightly from fluid to crystalline phases, with different close-packed stacking arrangements being distinguishable. In case of spherical particles $Q_6=0.4848$ for the hexagonal-close-packed structure and $Q_6=0.5745$ for face-centred-cubic structure, with other polytype phases having an order parameter value between these. As the dumbbells are not spherical and hence the in-plane and between-plane distances of the centres of masses will be different in a close-packed structure, the calculated $Q_6$ parameters are expected to deviate from the above values as the bong length increases.    
To characterise orientational order in the system, we define the nematic order parameter, $S$, as in Ref.~\cite{ghosh_model_1984}, 
$$
S=\langle P_{2}(\cos \theta )\rangle =\left\langle {\frac {3\cos ^{2}(\theta )-1}{2}}\right\rangle
$$
where $\theta$ is the angle between the chain and the mean direction vector of the configuration, i.e. the direction of the single bond in case of the dumbbell. The nematic order parameter is symmetric, not distinguishing the two spheres of the dumbbell, taking a value close to zero in orientationally disordered configurations and $S=1.0$ if all bonds point to the same direction. 

\begin{figure}[th]
    \centering    \includegraphics[angle=0,width=\linewidth]{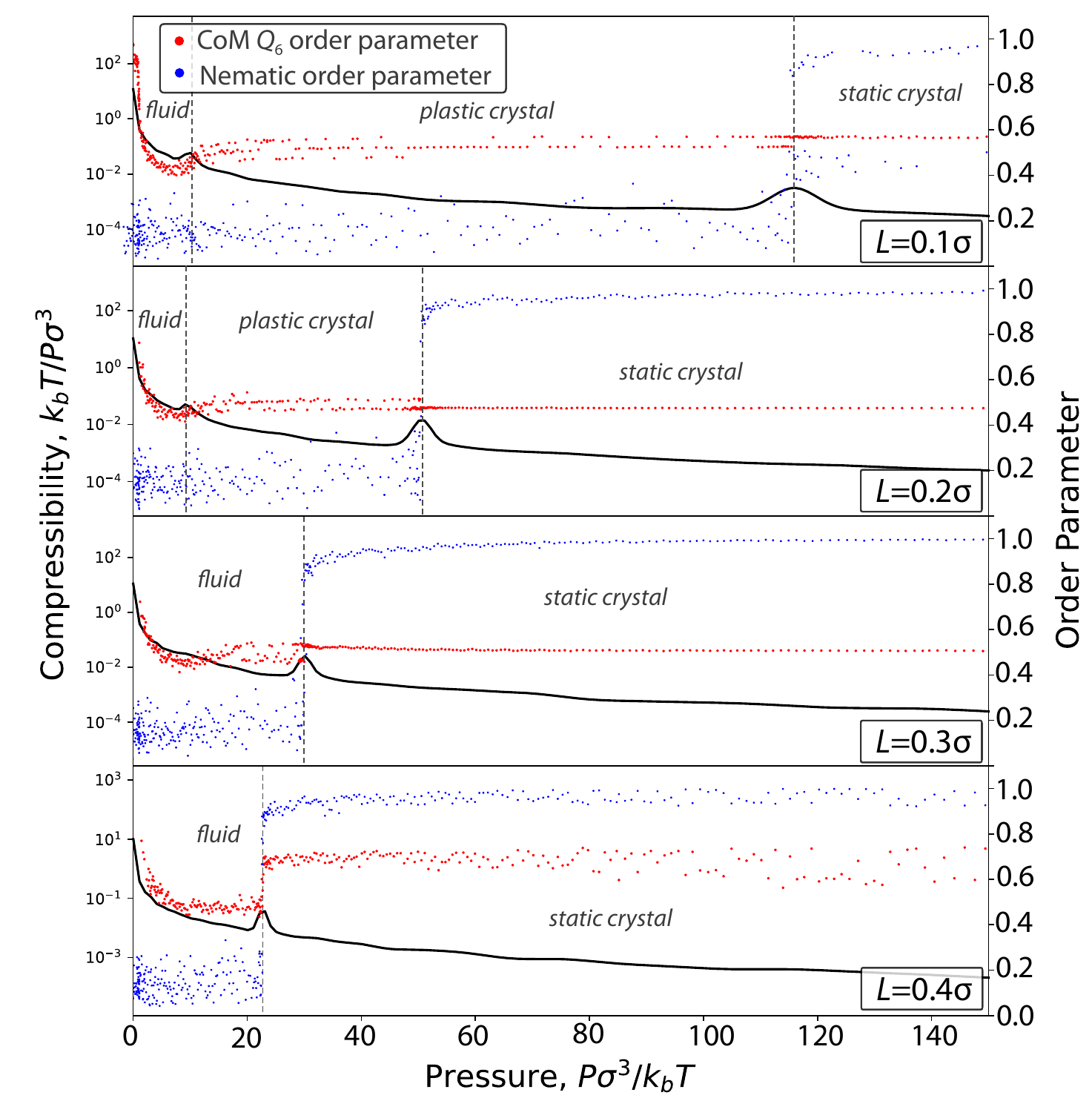}
    \caption{Compressibility (black solid line), average nematic order parameter (blue dots) and average $Q_6$ bond order parameter (red dots) of configurations generated during NS, plotted as a function of pressure, for the system of dumbbells of bond length 0.1 to 0.4$\sigma$. Vertical dashed lines highlight phase transitions.}
    \label{fig:nematic}
\end{figure}

\begin{figure}[htb]
    \begin{center}
    \includegraphics[width=\linewidth]{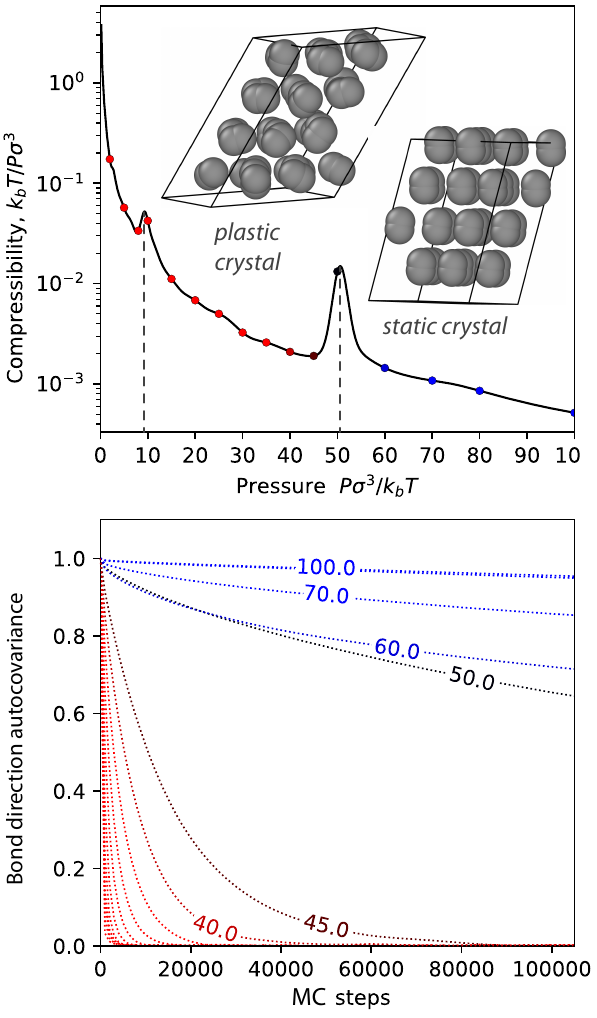}
    \end{center}
    \caption{Top panel: Compressibility curve calculated by nested sampling for $L=0.2\sigma$. Example solid configurations generated during the sampling are shown for the plastic crystal (dumbbells in ordered position but random orientation) and the ground state structure of a fully ordered static crystal. Spheres are shown smaller than their true size for clarity. Bottom panel: The mean autocovariances of the vector representing the bond over a MC run. The points on the compressibility curve correspond to the pressures at which the MC runs were performed (pressure values are also shown along some of the autocovariance lines.}
    \label{fig:bondcorrelation}
\end{figure}

In \autoref{fig:nematic} we show both order parameters calculated for configurations generated by nested sampling along with the compressibility curves for four different bond lengths. 
For $L=0.1$ and $0.2$ the lower pressure compressibility peak appears around $P^{*}=10$, a value similar to the fluid-solid transition observed in hard spheres \cite{hoover1968melting,noya2008determination,partay_nested_2014}. 
This transition is marked by a change in the $Q_6$ parameter, with at least two different values becoming dominant (corresponding to different stacking arrangements), suggesting that the dumbbells become arranged such that their centres of mass form a regular close-packed structure. 
However, the nematic order parameter remains unchanged and at a low value of $S<0.2$, meaning that the dumbbells are not orientationally aligned. 
To confirm whether this phase is a plastic crystal or an ensemble of glass-like jammed configurations \cite{torquato2000random,torquato2001multiplicity,partay_nested_2014}, we performed MC simulations and tracked the autocovariance of the bond directions. 
In a plastic crystal (or a fluid), where dumbbells can freely rotate and thus point to different directions as the simulation progresses, the autocovariance decays rapidly to zero. If the rotation of dumbbells is restricted, and thus their relative orientation changes very slowly or remains constant, such as in a jammed state or an orientationally ordered static crystal phase, the autocovariance does not decay or decays very slowly. 
\autoref{fig:bondcorrelation} shows the bond orientation autocovariance for $L=0.2$, with separate MC simulations of 256 particles performed at a range of different pressure values. These simulations were started from the static crystal structure, selecting a configuration generated by NS, and equilibrated via MC, maintaining a 50\% acceptance probability during the MC sweeps for each simulations. 
It is apparent that the bond orientation autocovariance decays very rapidly at lower pressures, even well beyond the low pressure compressibility peak, confirming that dumbbells can freely rotate, while their centres of mass occupies regular position, and hence this transition corresponds to a fluid-plastic crystal phase transition. This is not dissimilar to the plastic and rotator phases observed for example in  methane \cite{PhysRevB.55.14800}, ethane \cite{doi:10.1021/j100587a010} or recently in ice\cite{plastic_ice}, but is distinctly different from orientational glasses, such as those observed in low temperature KCN/KBr mixtures\cite{PhysRevB.22.1417}.

The location of this first phase transition on increasing $P$ is consistent with earlier literature\cite{vega_stability_1992,singer_monte_1990,marechal_stability_2008}.
The decay of the autocovariance only changes dramatically upon the pressure increasing to the second compressibility peak at higher pressure. 
We can see in \autoref{fig:nematic} that this is also marked by a sudden and significant increase in the nematic order parameter, showing that the dumbbells become orientationally aligned, pointing in the same direction. 
Thus, we can identify this second compressibility peak as a solid-solid transition between plastic and static crystal phases. 

\begin{figure}[h]
         \centering
         \includegraphics[width=\linewidth]{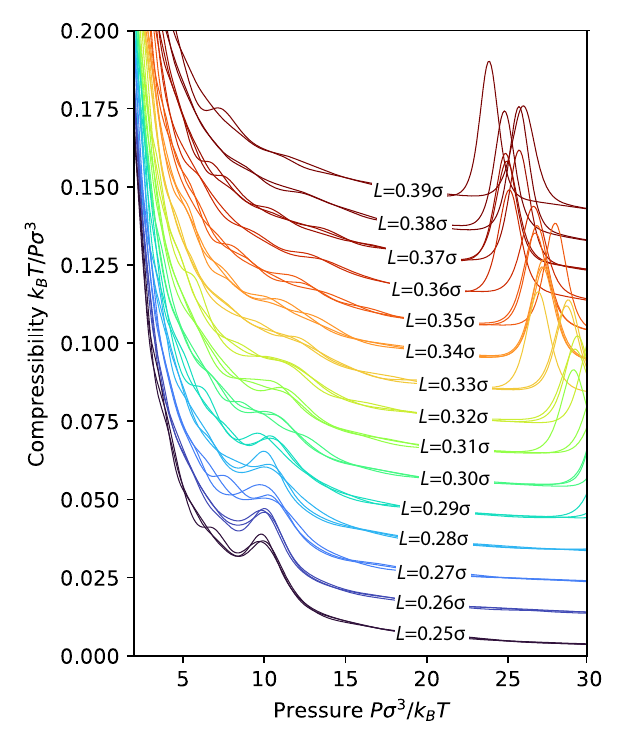}
         \caption{Compressibility as a function of pressure in the bond lengths range of $L=0.25-0.39$, showing the diminishing fluid-plastic crystal transition peak at pressures around $P^{*}=10$.}
         \label{fig:comp2}
\end{figure}

It is interesting to note that while the plastic to static crystal transition shifts to lower pressures as the bond length increases and thus the dumbbell becomes less spherical, the fluid to plastic transition pressure remains almost unchanged. 
However, we do not see evidence of the two transitions merging, as the compressibility peak corresponding to the fluid-plastic transition starts to diminish at $L=0.25$ and becomes unidentifiable by $L=0.32$, without any significant shift in the peak position, as seen in~\autoref{fig:comp2}.

\subsubsection{Bond lengths $0.4\sigma \leq L < 0.8\sigma$}
\label{sec:midbonds}

Beyond $L\approx 0.38$ prior studies suggest that our NS simulations should identify close packed crystal phases. However for bond lengths between $0.5< L< 0.8$, finding these crystalline structures proved to be challenging. Our NS simulations using 32 dumbbells often became trapped in amorphous jammed states, where translational and rotational moves no longer contribute to meaningful exploration of different structures. As noted earlier, no previous study has been able to capture nucleation of close packed CPx phases from the fluid \cite{ni_crystal_2011, zubieta_nucleation_2019}. The difficultly in reaching them with NS is consistent with those prior observations.
We can however still observe a transition peak from the fluid to these amorphous solid phases, albeit less pronounced and at slightly higher pressures than for the crystalline transition. 
The density of the amorphous structures is approximately 10\% lower than that of the thermodynamically stable crystal at the same pressure. 
Overall, this suggests that accessing the crystalline phase in these systems is difficult and would require the resolution of the sampling to be increased. However, doubling the number of walkers was still insufficient in these cases, and hence we decided to decrease the dimensionality of the problem by decreasing the system size.  
Thus, we performed NS calculations using 16 dumbbells, first with bond length $L=0.4$ to be able to compare the results to those generated with 32 dumbbells and therefore estimate the effect of the smaller size of the system on the location of the transition. We found that -- in agreement with previous studies \cite{baldock_determining_2016} -- using a smaller system causes the solid-fluid transition point to be overestimated: the transition pressure calculated with with 16 dumbbells is 8.3\% above the 32 dumbbell results, based on several independent calculations. 




Using 16 dumbbells and $K=1152$ walkers provided sufficient resolution to prevent the sampling from becoming trapped in a disordered jammed state, reaching the close packed  crystalline phase.

\begin{figure}
    \centering
    \includegraphics[width=0.9\linewidth]{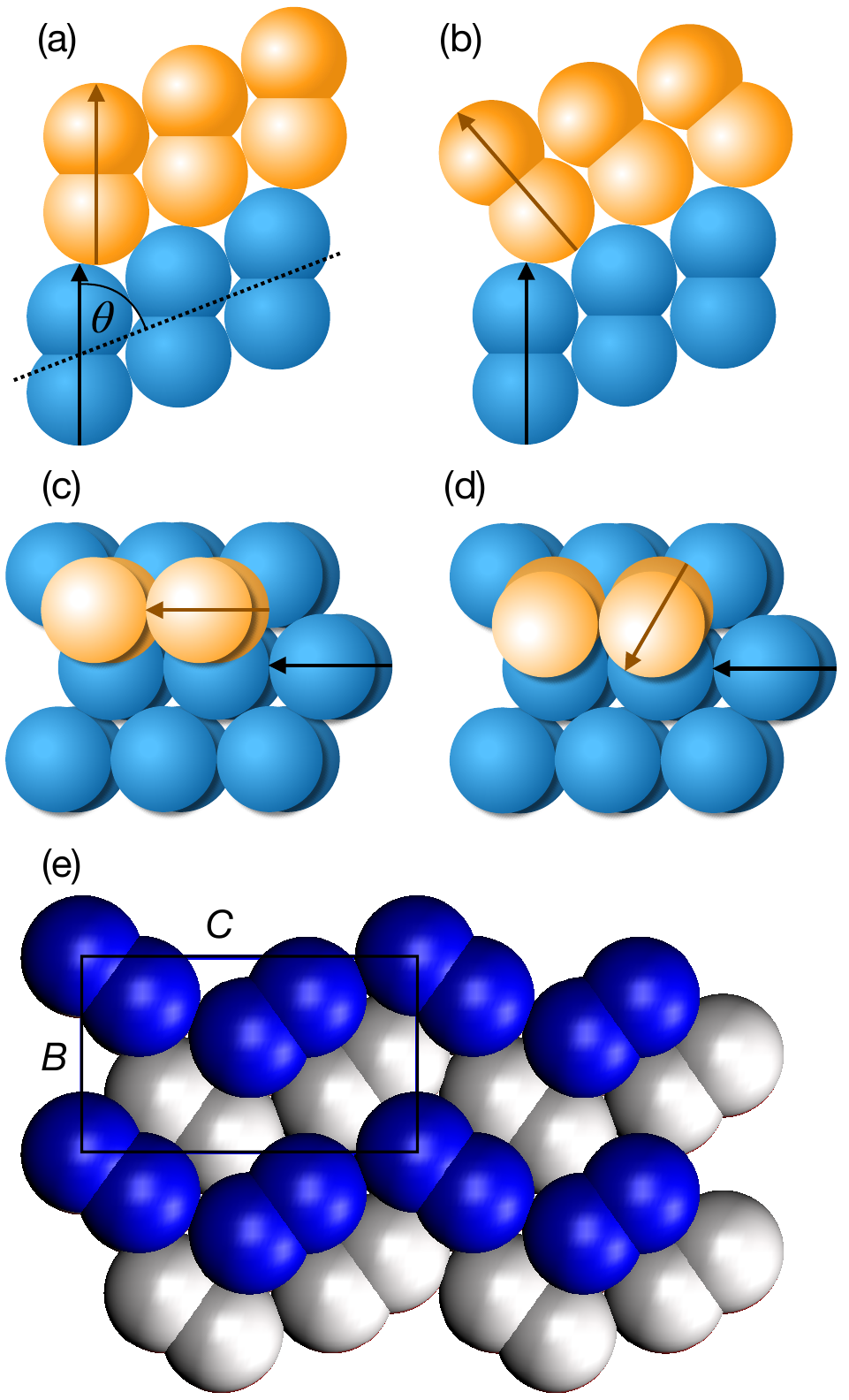}
    \caption{Panels (a)-(d): schematic representation of the close-packed stacked structure with various relative orientation of the dumbbells in consecutive layers. Arrows highlight the direction of the bond within a layer, $\theta$ is the tilt angle relative to the plane of the centre of the dumbbells. Panel (e): structure of the alternative packing with $Pnma$ symmetry for $L=0.6\sigma$, showing the orthogonal unit cell with four dumbbells. Colouring highlights dumbbells in the same plane. 
    }
    \label{fig:Pnma_crystal}
\end{figure}

As discussed earlier the possible arrangements of dumbbell into stacked close packed layers is richer than has been described previously in the literature. In addition to variations in layered stacking, the orientation and tilt of the dumbbells introduce further degrees of freedom as demonstrated in panels (a) to (d) of \autoref{fig:Pnma_crystal}.
The allowed degree of tilt ($\theta$ in \autoref{fig:Pnma_crystal} (a)) depends on the bond length as $L = 2\cos \theta$, since the densest packing is achieved when one of the spheres forming the dumbbell is positioned in the narrowest mid-section of a neighbouring dumbbell, making contact with both of its constituent spheres.
In case of a tilted layer, the direction of the tilt relative to the underlying layer allows further variations in the structure (\autoref{fig:Pnma_crystal} (c) and (d)). While these structures all have the same density at maximum packing, their thermodynamic properties, such as the entropy, will differ due to differences in the second and subsequent neighbour shells. Example structures are presented in supplementary information section \ref{sec:CPx}. Several of these were observed in our NS simulations, many more than the three examples CP[1-3] considered previously in the literature. 

While the close packed crystalline structures sampled by NS remained largely similar across the bond length range $0.1\leq L \leq0.8$ the distribution of dumbbell orientations appeared to increase with bond length. This potentially indicates that stackings such as CP1 and CP2 in which all dumbbells share the same orientation are marginally less stable than others at larger $L$. This will require detailed free energy calculations to confirm, which we defer to future work. We note however than a study by \citeauthor{kowalik_free_2008} \cite{kowalik_free_2008} found the CP3 structure (in which dumbbell orientations alternate between layers) to have equal free energy to CP1 and CP2 at $L=1$ within the precision of their calculations. 

Interestingly, an entirely different class of close packed crystalline phases also emerged from the NS simulations. These cannot be considered as stacked layers of close packed dumbbells but have the same maximum packing fraction as the CPx  structures for any bond length. As an example we consider a structure with $Pnma$ symmetry. The orthogonal unit cell of this structure contains two pairs of dumbbells. The four spheres of each pair lie in a plane, with their bonds forming the tetrahedral angle of $\approx109.5^\circ$, and the two planes being of $\sigma/2$ distance from each other. Each dumbbell is parallel to one other within the unit cell. At the limit of $L=0.0$, the structure reduces to the fcc structure, while in case of $L=1.0$, the individual spheres display a double-hexagonal-close-packed (dhcp) arrangement. At $L=1$ this structure may be considered as one particular realisation of the 3D aperiodic structure with ordered bonds, but is distinct from the equivalent CPx structures. Specifically one sphere in the 2D rectangular unit cell of the close packed layer is bonded to a sphere in the layer above, and one to the layer below. Bonds are hence interdigitated between successive bilayers. At smaller $L$ the \emph{positions} of the spheres within the unit cell are also different from the equivalent CPx structure.  This structure is shown in panel (e) of \autoref{fig:Pnma_crystal} for $L=0.6$.
The dimensions of the unit cell, assuming the diameter of the individual hard spheres is 1.0, are $B=\frac{2\sqrt{3+L^2+2L\sqrt{6-2L^2}}}{\sqrt{3}}$, $C=\frac{\sqrt{6-L^2+2L\sqrt{6-2L^2}}}{\sqrt{3}}$ and for the third axis perpendicular to the projection shown in the figure, $A=1.0$. An alternative description of this structure and its possible variants (by analogy with the CPx structures) is presented in the supplementary information section \ref{sec:CPx-interdig}. In addition, in supplementary information section \ref{sec:CPpf} we present results that indicate these structures can be slightly higher in density than the previously known structures when simulated at pressures just above the crystallisation pressure.


\subsubsection{Bond lengths $0.8\sigma < L \leq 1.0\sigma$}

\begin{figure}[ht]
    \centering
    \includegraphics[width=\linewidth]{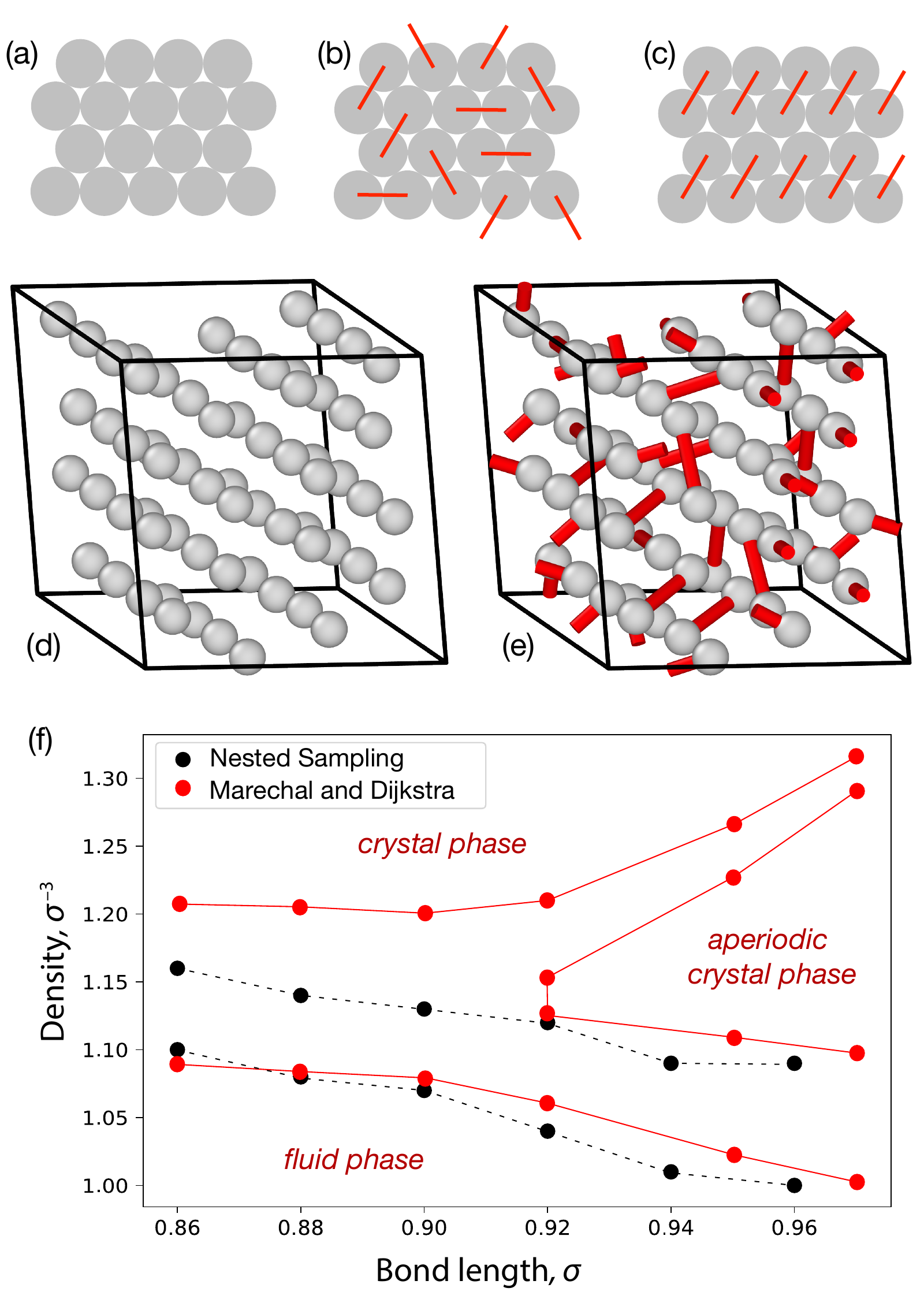}
    \caption{Representation of the aperiodic crystal structure with $L=1.0\sigma$. Schematic drawings represent structures of the same packing fraction of (a) close packed circles in 2D (b) aperiodic crystal of dumbbells in 2D (c) perfectly aligned dumbbells in 2D. Panels (d) and (e) show a configuration from the end of a nested sampling run with $L=1.0\sigma$, without and with the bonds shown by red lines, respectively. Spheres are shown by size smaller then $\sigma$ to make bonds visible. Panel (f) shows the bond-length--density phase diagram, comparing NS results and those from Ref. \cite{marechal_stability_2008}.}
     \label{fig:aperiodic}
\end{figure}

At longer bond lengths of $L\geq0.8$, the distance between the two hard spheres composing the dumbbell becomes comparable to the diameter of the individual spheres. 
This means that the overall shape of a dumbbell closely resembles two spheres with minimal overlap, and in the limit of $L = 1.0$, the highest density dumbbell packing will become identical to that of single hard spheres. However, such a phase is realisable with different orientational arrangements of the dumbbells as shown in \autoref{fig:aperiodic}. 
As a result, a new solid phase become distinguishable from the orientationally ordered crystal, where the spacial arrangement of the dumbbells do not display translational symmetry. This phase is often referred to as the \emph{aperiodic crystal} and has been shown to be thermodynamically stable~\cite{vega2001extending,marechal_stability_2008}.

We find this aperiodic phase emerging in the nested sampling simulations, when $L\geq0.86$. 
\autoref{fig:aperiodic} shows this region of the bond-length--density phase diagram, highlighting the three thermodynamically stable phases and the coexistence regions in-between, comparing our result to MC simulations from Ref.~\cite{marechal_stability_2008}
In case of the NS simulations, the density of the phase boundaries were determined by calculating the average density at the pressure values before and after the corresponding compressibility peak.
The resulting phase boundaries of both the fluid and the aperiodic phases agree well with previous findings, however, our sampling does not capture the solid-solid transition to the fully ordered crystal phase.
As the entropy of the orientationally ordered phase must be substantially lower than that of the aperiodic crystal, we can assume that a sampling with significantly higher resolution would be necessary to sample both basins and hence observe the transition with nested sampling. 

\section{Conclusion}

We performed an extensive sampling of the configuration space of hard dumbbells of various values of bond lengths in the range of $L=0.1-1.0$. 
We used Monte Carlo simulations as well as the nested sampling method to locate phase transitions and identify thermodynamically stable phases.

\begin{figure}[hb]
    \centering
    \includegraphics[width=0.5\textwidth]{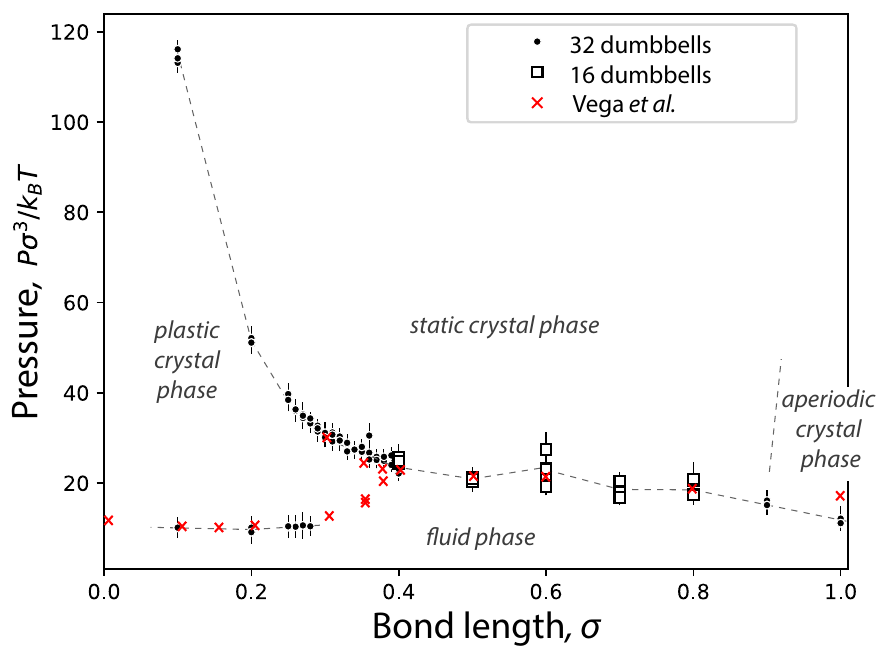}
    \centering
   \caption{Bond-length--pressure phase diagram of hard dumbbells, showing the thermodynamically stable phases. Black circles and open squares show phase transition pressures determined from nested sampling calculations of the current work, using 32 and 16 dumbbells, respectively. Red crosses show results from Ref.~\cite{vega_stability_1992}. Dashed lines are only guides to the eye.} \label{fig:Phase_Diag_1}
\end{figure}

\autoref{fig:Phase_Diag_1} summarises our results across the entire range of studied dumbbells in the form of a bond-length--pressure phase diagram, also showing the Monte Carlo simulation results obtained by Vega et al. \cite{vega_stability_1992}, for comparison.
We identified four thermodynamically stable phases: fluid phase at low pressures, plastic crystal phase with dumbbells occupying crystalline lattice sites but rotating freely at short bond lengths, with this phase undergoing a further transition and becoming fully aligned at high pressure and longer bond length, as well as a fourth phase of aperiodic crystal, where neither the orientation, nor the centres of mass of the dumbbells show long range order. 

Apart from the excellent agreement with existing results in the literature, our findings demonstrate how nested sampling can enhance our understanding of phase stability and lead to the discovery of new crystalline structures. Beyond providing a more detailed characterization of the plastic crystal–static crystal transition and also highlighting the diverse dumbbell orientations in the static crystalline phase, we identified an entirely new family of close-packed crystalline structures which have identical, or slightly higher packing fractions than those considered in previous studies. This underscores the importance of unbiased sampling in uncovering (novel) packing patterns.

Crucially, this work represents the first systematic application of nested sampling to non-spherical particles resembling molecules. While softer potentials are less prone to challenges associated with jammed structures, our results indicate that the added rotational degrees of freedom may require higher-resolution sampling than what has been found sufficient in applications to atomic materials.


\section*{Data availability}
The data supporting this study's findings will be made available at the publication stage. Software to perform NS and MC calculations as well as to produce different close-packed structures are available on GitHub at \url{https://github.com/omaradesida/hans}, \url{https://github.com/dquigley533/hs_alkane} and \url{https://github.com/dquigley533/Hard-Dumbbell-CP/tree/main}, respectively.

\section*{Acknowledgements}
The Authors thank Gordon Bart\'ok-P\'artay for his contributions to the preliminary formulae and the initial code to generate coordinates of the new packing structure.
O.A. acknowledges funding from the EPSRC Centre for Doctoral Training in Modelling of Heterogeneous Systems (EP/S022848/1).
L.B.P. acknowledges support from the EPSRC through an individual Early Career Fellowship (EP/T000163/1). 
Computing facilities were provided by the Scientific Computing Research Technology Platform of the University of Warwick. 
Part of the calculations were performed using the Sulis Tier 2 HPC platform hosted by the
Scientific Computing Research Technology Platform at the University of Warwick.
Sulis is funded by EPSRC Grant EP/T022108/1 and the HPC Midlands+ consortium.

\bibliography{Dimer}

@article{marechal_stability_2008,
	title = {Stability of orientationally disordered crystal structures of colloidal hard dumbbells},
	volume = {77},
	issn = {1539-3755, 1550-2376},
	url = {https://link.aps.org/doi/10.1103/PhysRevE.77.061405},
	doi = {10.1103/PhysRevE.77.061405},
	pages = {061405},
	number = {6},
	journal = {Phys. Rev. E},
	author = {Marechal, Matthieu and Dijkstra, Marjolein},
	urldate = {2021-12-01},
	year = {2008},
	langid = {english},
}

@article{vega_plastic_1997,
	title = {Plastic crystal phases of hard dumbbells and hard spherocylinders},
	volume = {107},
	issn = {0021-9606, 1089-7690},
	url = {http://aip.scitation.org/doi/10.1063/1.474626},
	doi = {10.1063/1.474626},
	pages = {2696--2697},
	number = {7},
	journal = {J. Chem. Phys.},
	author = {Vega, C. and Monson, P. A.},
	urldate = {2021-12-01},
	year = {1997},
	langid = {english},
}

@article{noya2008determination,
  title={Determination of the melting point of hard spheres from direct coexistence simulation methods},
  author={Noya, Eva G and Vega, Carlos and de Miguel, Enrique},
  journal={J. Chem. Phys.},
  url={https://doi.org/10.1063/1.2901172},
  volume={128},
  pages={154507},
  number={15},
  year={2008},
  publisher={AIP Publishing}
}

@software{quigley_hs_alkane_2019,
	title = {hs\_alkane},
	url = {https://github.com/dquigley533/hs_alkane},
	version = {0.0.0},
	author = {Quigley, David},
	urldate = {2023-08-31},
	year = {2019},
	note = {original-date: 2018-08-16T14:55:12Z},
}

@article{partay_nested_2021,
	title = {Nested sampling for materials},
	volume = {94},
	issn = {1434-6036},
	url = {https://doi.org/10.1140/epjb/s10051-021-00172-1},
	doi = {10.1140/epjb/s10051-021-00172-1},
	pages = {159},
	number = {8},
	journal = {Eur. Phys. J. B},
	author = {Pártay, Livia B. and Csányi, Gábor and Bernstein, Noam},
	urldate = {2023-08-31},
	year = {2021},
	langid = {english},
}

@article{alder_phase_1962,
	title = {Phase Transition in Elastic Disks},
	volume = {127},
	issn = {0031-899X},
	url = {https://link.aps.org/doi/10.1103/PhysRev.127.359},
	doi = {10.1103/PhysRev.127.359},
	pages = {359--361},
	number = {2},
	journal = {Phys. Rev.},
	author = {Alder, B. J. and Wainwright, T. E.},
	urldate = {2023-11-10},
	year = {1962},
	langid = {english},
}

@article{Tildesley01091980,
author = {D.J. Tildesley and W.B. Streett},
title = {An equation of state for hard dumbell fluids},
journal = {Mol. Phys.},
volume = {41},
number = {1},
pages = {85--94},
year = {1980},
publisher = {Taylor \& Francis},
URL = { https://doi.org/10.1080/00268978000102591},
}

@article{malanoski_solid-fluid_1999,
	title = {Solid-fluid equilibrium in molecular models of n-alkanes},
	volume = {110},
	issn = {0021-9606},
	url = {https://doi.org/10.1063/1.478123},
	doi = {10.1063/1.478123},
	pages = {664--675},
	number = {1},
	journal = {J. Chem. Phys.},
	author = {Malanoski, A. P. and Monson, P. A.},
	urldate = {2021-09-08},
	year = {1999},
}

@article{bridgwater_lattice-switching_2014,
	title = {Lattice-switching Monte Carlo method for crystals of flexible molecules},
	volume = {90},
	url = {https://link.aps.org/doi/10.1103/PhysRevE.90.063313},
	doi = {10.1103/PhysRevE.90.063313},
	pages = {063313},
	number = {6},
	journal = {Phys. Rev. E},
	author = {Bridgwater, Sally and Quigley, David},
	year = {2014},
}

@article{doi:10.1021/j100587a010,
author = {Eggers, D. F. Jr.},
title = {Plastic-crystalline phase of ethane},
journal = {J. Phys. Chem.},
volume = {79},
number = {20},
pages = {2116-2118},
year = {1975},
URL = { https://doi.org/10.1021/j100587a010},
}

@article{PhysRevB.55.14800,
  title = {High-pressure infrared study of solid methane: Phase diagram up to 30 GPa},
  author = {Bini, Roberto and Pratesi, Gabriele},
  journal = {Phys. Rev. B},
  volume = {55},
  issue = {22},
  pages = {14800--14809},
  numpages = {0},
  year = {1997},
  month = {Jun},
  publisher = {American Physical Society},
  doi = {10.1103/PhysRevB.55.14800},
  url = {https://0-link-aps-org.pugwash.lib.warwick.ac.uk/doi/10.1103/PhysRevB.55.14800}
}

@article{zubieta_nucleation_2019,
	title = {Nucleation of pseudo hard-spheres and dumbbells at moderate metastability: appearance of A15 Frank–Kasper phase at intermediate elongations},
	volume = {21},
	issn = {1463-9084},
	url = {http://0.pubs.rsc.org/en/content/articlelanding/2019/cp/c8cp04964e},
	doi = {10.1039/C8CP04964E},
	shorttitle = {Nucleation of pseudo hard-spheres and dumbbells at moderate metastability},
	pages = {1656--1670},
	number = {4},
	journal = {Phys. Chem. Chem. Phys.},
	author = {Zubieta, Itziar and Saz, Miguel Vázquez del and Llombart, Pablo and Vega, Carlos and Noya, Eva G.},
	urldate = {2022-02-08},
	year = {2019},
}

@article{vega_solidfluid_1992,
	title = {Solid–fluid equilibria for hard dumbbells via Monte Carlo simulation},
	volume = {96},
	issn = {0021-9606, 1089-7690},
	url = {http://aip.scitation.org/doi/10.1063/1.462214},
	doi = {10.1063/1.462214},
	pages = {9060--9072},
	number = {12},
	journal = {J. Chem. Phys.},
	author = {Vega, C. and Paras, E. P. A. and Monson, P. A.},
	urldate = {2022-02-08},
	year = {1992},
}

@article{vega_stability_1992,
	title = {On the stability of the plastic crystal phase of hard dumbbell solids},
	volume = {97},
	issn = {0021-9606, 1089-7690},
	url = {http://aip.scitation.org/doi/10.1063/1.463372},
	doi = {10.1063/1.463372},
	pages = {8543--8548},
	number = {11},
	journal = {J. Chem. Phys.},
	author = {Vega, C. and Paras, E. P. A. and Monson, P. A.},
	urldate = {2022-02-09},
	year = {1992},
}

@article{ni_crystal_2011,
	title = {Crystal nucleation of colloidal hard dumbbells},
	volume = {134},
	issn = {0021-9606, 1089-7690},
	url = {http://aip.scitation.org/doi/10.1063/1.3528222},
	doi = {10.1063/1.3528222},
	pages = {034501},
	number = {3},
	journal = {J. Chem. Phys.},
	author = {Ni, Ran and Dijkstra, Marjolein},
	urldate = {2022-05-27},
	year = {2011},
}

@article{skilling_nested_2006,
	title = {Nested sampling for general Bayesian computation},
	volume = {1},
	issn = {1936-0975, 1931-6690},
	url = {https://projecteuclid.org/journals/bayesian-analysis/volume-1/issue-4/Nested-sampling-for-general-Bayesian-computation/10.1214/06-BA127.full},
	doi = {10.1214/06-BA127},
	pages = {833--859},
	number = {4},
	journal = {Bayesian Analysis},
	author = {Skilling, John},
	urldate = {2022-11-01},
	year = {2006},
}

@article{ashton_nested_2022,
	title = {Nested sampling for physical scientists},
	volume = {2},
	issn = {2662-8449},
	url = {https://www.nature.com/articles/s43586-022-00121-x},
	doi = {10.1038/s43586-022-00121-x},
	pages = {39},
	number = {1},
	journal = {Nat Rev Methods Primers},
	author = {Ashton, Greg and Bernstein, Noam and Buchner, Johannes and Chen, Xi and Csányi, Gábor and Fowlie, Andrew and Feroz, Farhan and Griffiths, Matthew and Handley, Will and Habeck, Michael and Higson, Edward and Hobson, Michael and Lasenby, Anthony and Parkinson, David and Pártay, Livia B. and Pitkin, Matthew and Schneider, Doris and Speagle, Joshua S. and South, Leah and Veitch, John and Wacker, Philipp and Wales, David J. and Yallup, David},
	urldate = {2022-11-01},
	year = {2022},
}

@article{grigoriev_monte-carlo_2020,
	title = {Monte-Carlo determination of adiabatic compressibility of hard spheres},
	volume = {46},
	issn = {0892-7022},
	url = {https://doi.org/10.1080/08927022.2020.1789124},
	doi = {10.1080/08927022.2020.1789124},
	pages = {905--910},
	number = {12},
	journal = {Mol. Sim.},
	author = {Grigoriev, A. N. and Kleshchonok, T. V. and Markov, I. V. and Bulavin, L. A.},
	urldate = {2022-11-01},
	year = {2020},
}

@article{dorrell_pressuretemperature_2020,
	title = {Pressure–Temperature Phase Diagram of Lithium, Predicted by Embedded Atom Model Potentials},
	volume = {124},
	pages = {6015--6023},
    url = {https://pubs.acs.org/doi/10.1021/acs.jpcb.0c03882},
	journal = {J. Phys. Chem. B},
	author = {Dorrell, Jordan and Pártay, Livia B.},
	year = {2020},
}

@article{bartok_insight_2021,
	title = {Insight into Liquid Polymorphism from the Complex Phase Behavior of a Simple Model},
	volume = {127},
	pages = {015701},
    url = {https://doi.org/10.1103/PhysRevLett.127.015701},
	journal = {Phys. Rev. Lett.},
	author = {Bartók, Albert P. and Hantal, György and Pártay, Livia B.},
	year = {2021},
}

@article{baldock_constant-pressure_2017,
	title = {Constant-pressure nested sampling with atomistic dynamics},
	volume = {96},
	pages = {43311--43324},
	journal = {Phys. Rev. E},
    url = {https://doi.org/10.1103/PhysRevE.96.043311},
	author = {Baldock, R. J. N. and Bernstein, N. and Salerno, K. M. and Pártay, L. B. and Csányi, G.},
	year = {2017},
}

@article{partay_nested_2014,
	title = {Nested sampling for materials: The case of hard spheres},
	volume = {89},
	pages = {022302},
    url = {https://doi.org/10.1103/PhysRevE.89.022302},
	journal = {Phys. Rev. E},
	author = {Pártay, L. B. and Bartók, A. P. and Csányi, G.},
	year = {2014},
}

@article{baldock_determining_2016,
	title = {Determining pressure-temperature phase diagrams of materials},
	volume = {93},
	url = {http://link.aps.org/doi/10.1103/PhysRevB.93.174108},
	doi = {10.1103/PhysRevB.93.174108},
	pages = {174108},
	number = {17},
	journal = {Phys. Rev. B},
	author = {Baldock, Robert J. N. and Pártay, Lívia B. and Bartók, Albert P. and Payne, Michael C. and Csányi, Gábor},
	year = {2016},
}

@article{partay_efficient_2010,
	title = {Efficient Sampling of Atomic Configurational Spaces},
	volume = {114},
	pages = {10502--10512},
	journal = {J. Phys. Chem. B},
	author = {Pártay, L. B. and Bartók, A. P. and Csányi, G.},
    url={https://pubs.acs.org/doi/10.1021/jp1012973},
	year = {2010},
}

@article{ghosh_model_1984,
	title = {A model for the orientational order in liquid crystals},
	volume = {4},
	url = {https://0-link-springer-com.pugwash.lib.warwick.ac.uk/article/10.1007/BF02453342},
	doi = {10.1007/BF02453342},
	pages = {229--244},
	number = {3},
	journal = {Il Nuovo Cimento D},
	author = {Ghosh, S. K.},
	urldate = {2023-03-03},
	year = {1984},
}

@article{alder_phase_1957,
	title = {Phase Transition for a Hard Sphere System},
	volume = {27},
	issn = {0021-9606, 1089-7690},
	url = {https://pubs.aip.org/jcp/article/27/5/1208/204728/Phase-Transition-for-a-Hard-Sphere-System},
	doi = {10.1063/1.1743957},
	pages = {1208--1209},
	number = {5},
	journal = {J. Chem. Phys.},
	author = {Alder, B. J. and Wainwright, T. E.},
	urldate = {2023-11-13},
	year = {1957},
	langid = {english},
}

@article{ohern_jamming_2003,
	title = {Jamming at zero temperature and zero applied stress: The epitome of disorder},
	volume = {68},
	issn = {1063-651X, 1095-3787},
	url = {https://link.aps.org/doi/10.1103/PhysRevE.68.011306},
	doi = {10.1103/PhysRevE.68.011306},
	shorttitle = {Jamming at zero temperature and zero applied stress},
	pages = {011306},
	number = {1},
	journal = {Phys. Rev. E},
	author = {O’Hern, Corey S. and Silbert, Leonardo E. and Liu, Andrea J. and Nagel, Sidney R.},
	urldate = {2023-11-13},
	year = {2003},
	langid = {english},
}

@article{vega_linear_1994,
	title = {Linear hard sphere models Virial coefficients and equation of state},
	volume = {82},
	issn = {0026-8976},
	url = {https://doi.org/10.1080/00268979400100874},
	doi = {10.1080/00268979400100874},
	pages = {1233--1247},
	number = {6},
	journal = {Mol. Phys.},
	author = {Vega, Carlos and Lago, Santiago and Garzón, Benito},
	urldate = {2023-11-13},
	year = {1994},
}

@article{boublik_equation_1977,
	title = {Equation of state for hard dumbbells},
	volume = {46},
	issn = {00092614},
	url = {https://linkinghub.elsevier.com/retrieve/pii/000926147785269X},
	doi = {10.1016/0009-2614(77)85269-X},
	pages = {315--316},
	number = {2},
	journal = {Chem. Phys. Lett.},
	author = {Boublík, T. and Nezbeda, I.},
	urldate = {2023-11-13},
	year = {1977},
	langid = {english},
}

@article{whittle_liquid_1991,
	title = {Liquid crystal formation in a system of fused hard spheres},
	volume = {72},
	issn = {0026-8976},
	url = {https://doi.org/10.1080/00268979100100191},
	doi = {10.1080/00268979100100191},
	pages = {247--265},
	number = {2},
	journal = {Mol. Phys.},
	author = {Whittle, M. and Masters, A.J.},
	urldate = {2023-11-13},
	year = {1991},
}

@article{charbonneau_memory_2021,
	title = {Memory Formation in Jammed Hard Spheres},
	volume = {126},
	url = {https://link.aps.org/doi/10.1103/PhysRevLett.126.088001},
	doi = {10.1103/PhysRevLett.126.088001},
	pages = {088001},
	number = {8},
	journal = {Phys. Rev. Lett.},
	author = {Charbonneau, Patrick and Morse, Peter K.},
	urldate = {2023-11-16},
	year = {2021},
}

@article{tasios_glassy_2014,
	title = {Glassy dynamics of convex polyhedra},
	volume = {141},
	issn = {0021-9606},
	url = {https://doi.org/10.1063/1.4902992},
	doi = {10.1063/1.4902992},
	pages = {224502},
	number = {22},
	journal = {J. Chem. Phys.},
	author = {Tasios, Nikos and Gantapara, Anjan Prasad and Dijkstra, Marjolein},
	year = {2014},
}

@article{gola_embedded_2018,
	title = {Embedded atom method potential for studying mechanical properties of binary Cu–Au alloys},
	volume = {26},
	issn = {0965-0393},
	url = {https://dx.doi.org/10.1088/1361-651X/aabce4},
	doi = {10.1088/1361-651X/aabce4},
	pages = {055006},
	number = {5},
	journal = {Modelling Simul. Mater. Sci. Eng.},
	author = {Gola, Adrien and Pastewka, Lars},
	urldate = {2023-11-17},
	year = {2018},
	langid = {english},
}

@article{radu_solid-solid_2009,
	title = {Solid-solid phase transition in hard ellipsoids},
	volume = {131},
	issn = {0021-9606, 1089-7690},
	url = {https://pubs.aip.org/jcp/article/131/16/164513/71316/Solid-solid-phase-transition-in-hard-ellipsoids},
	doi = {10.1063/1.3251054},
	pages = {164513},
	number = {16},
	journal = {J. Chem. Phys.},
	author = {Radu, M. and Pfleiderer, P. and Schilling, T.},
	urldate = {2023-11-17},
	year = {2009},
	langid = {english},
}

@article{pusey_hard_2009,
	title = {Hard spheres: crystallibtion and glass formation},
	volume = {367},
	url = {https://0-royalsocietypublishing-org.pugwash.lib.warwick.ac.uk/doi/full/10.1098/rsta.2009.0181},
	doi = {10.1098/rsta.2009.0181},
	shorttitle = {Hard spheres},
	pages = {4993--5011},
	number = {1909},
	journal = {Phil. Trans. R. Soc. A-Math. Phys. Eng. Sci.},
	author = {Pusey, P. N. and Zaccarelli, E. and Valeriani, C. and Sanz, E. and Poon, Wilson C. K. and Cates, Michael E.},
	urldate = {2023-11-17},
	year = {2009},
}

@software{adesida_hans_2023,
	title = {hans},
	url = {https://github.com/omaradesida/hans},
	author = {Adesida, Omar},
	urldate = {2023-11-20},
	year = {2023},
	note = {original-date: 2021-10-04T13:58:14Z},
}

@article{davidchack_direct_2000,
	title = {Direct Calculation of the Hard-Sphere Crystal {\textless}span class="aps-inline-formula"{\textgreater}{\textless}math xmlns="http://www.w3.org/1998/Math/{MathML}" display="inline"{\textgreater}{\textless}mi{\textgreater}/{\textless}/mi{\textgreater}{\textless}/math{\textgreater}{\textless}/span{\textgreater}Melt Interfacial Free Energy},
	volume = {85},
	doi = {10.1103/PhysRevLett.85.4751},
	shorttitle = {Direct Calculation of the Hard-Sphere Crystal {\textless}span class="aps-inline-formula"{\textgreater}{\textless}math xmlns="http},
	pages = {4751--4754},
	number = {22},
	journal = {Phys. Rev. Lett.},
	author = {Davidchack, Ruslan L.},
	year = {2000},
}

@article{QW,
        author = "P. J. Steinhardt and D. R. Nelson and M. Ronchetti",
        year = "1983",
        journal = "Phys. Rev. B",
        title = "Bond-orientational order in liquids and glasses",
        url = "https://doi.org/10.1103/PhysRevB.28.784",
        volume = "28",
        pages = "784"
}

@article{avendano2012phase,
  title={Phase behavior of rounded hard-squares},
  author={Avenda\~no, Carlos and Escobedo, Fernando A},
  journal={Soft Matter},
  volume={8},
  number={17},
  pages={4675--4681},
  year={2012},
  url={https://pubs.rsc.org/en/content/articlelanding/2012/sm/c2sm07428a},
  publisher={Royal Society of Chemistry}
}

@article{avendano2017packing,
  title={Packing, entropic patchiness, and self-assembly of non-convex colloidal particles: A simulation perspective},
  author={Avenda{\~n}o, Carlos and Escobedo, Fernando A},
  journal={Current opinion in colloid \& interface science},
  volume={30},
  url={https://doi.org/10.1016/j.cocis.2017.05.005},
  pages={62--69},
  year={2017},
  publisher={Elsevier}
}

@article{royall2023colloidal,
  title={Colloidal hard spheres: Triumphs, challenges and mysteries},
  author={Royall, C Patrick and Charbonneau, Patrick and Dijkstra, Marjolein and Russo, John and Smallenburg, Frank and Speck, Thomas and Valeriani, Chantal},
  journal={arXiv preprint arXiv:2305.02452},
  year={2023}
}

@article{miller2010phase,
  title={On the phase behavior of hard aspherical particles},
  author={Miller, William L and Cacciuto, Angelo},
  journal={J. Chem. Phys.},
  volume={133},
  number={23},
  year={2010},
  url={https://doi.org/10.1063/1.3518976},
  publisher={AIP Publishing}
}

@article{anderson2020hoomd,
  title={HOOMD-blue: A Python package for high-performance molecular dynamics and hard particle Monte Carlo simulations},
  author={Anderson, Joshua A and Glaser, Jens and Glotzer, Sharon C},
  journal={Comput. Mater. Sci.},
  volume={173},
  pages={109363},
  url={https://doi.org/10.1016/j.commatsci.2019.109363},
  year={2020},
  publisher={Elsevier}
}

@article{hoover1968melting,
  title={Melting transition and communal entropy for hard spheres},
  author={Hoover, William G and Ree, Francis H},
  journal={J. Chem. Phys.},
  volume={49},
  url={https://doi.org/10.1063/1.1670641},
  number={8},
  pages={3609--3617},
  year={1968},
  publisher={American Institute of Physics}
}

@article{torquato2000random,
  title={Is random close packing of spheres well defined?},
  author={Torquato, Salvatore and Truskett, Thomas M and Debenedetti, Pablo G},
  journal={Phys. Rev. Lett.},
  volume={84},
  number={10},
  url={https://doi.org/10.1103/PhysRevLett.84.2064},
  pages={2064},
  year={2000},
  publisher={APS}
}

@article{torquato2001multiplicity,
  title={Multiplicity of generation, selection, and classification procedures for jammed hard-particle packings},
  author={Torquato, Salvatore and Stillinger, Frank H},
  journal={J. Phys. Chem. B},
  volume={105},
  number={47},
  url={https://pubs.acs.org/doi/10.1021/jp011960q},
  pages={11849--11853},
  year={2001},
  publisher={ACS Publications}
}

@article{vega2001extending,
  title={Extending Wertheim’s perturbation theory to the solid phase: The freezing of the pearl-necklace model},
  author={Vega, C and MacDowell, LG},
  journal={J. Chem. Phys.},
  volume={114},
  url={https://doi.org/10.1063/1.1372329},
  number={23},
  pages={10411--10418},
  year={2001},
  publisher={American Institute of Physics}
}

@article{PhysRevB.22.1417,
  title = {Existence of an orientational glass state in ${(\mathrm{KCN})}_{x}{(\mathrm{KBr})}_{1\ensuremath{-}x}$ mixed crystals},
  author = {Michel, K. H. and Rowe, J. M.},
  journal = {Phys. Rev. B},
  volume = {22},
  issue = {3},
  pages = {1417--1428},
  numpages = {0},
  year = {1980},
  month = {Aug},
  publisher = {American Physical Society},
  doi = {10.1103/PhysRevB.22.1417},
  url = {https://link.aps.org/doi/10.1103/PhysRevB.22.1417}
}

@article{plastic_ice,
  title={Observation of plastic ice VII by quasi-elastic neutron scattering},
  author={Rescigno, Maria and Toffano, Alberto and Ranieri, Umbertoluca and Andriambariarijaona, Leon and Gaal, Richard and Klotz, Stefan and Koza, Michael Marek and Ollivier, Jacques and Martelli, Fausto and Russo, John and  Francesco Sciortino and Jose Teixeira and Livia Eleonora Bove},
  journal={Nature},
  pages={662–667},
  volume={640},
  year={2025},
  url={https://www.nature.com/articles/s41586-025-08750-4},
  publisher={Nature Publishing Group UK London}
}

@article{PhysRevE.61.906,
  title = {Lattice-switch Monte Carlo method},
  author = {Bruce, A. D. and Jackson, A. N. and Ackland, G. J. and Wilding, N. B.},
  journal = {Phys. Rev. E},
  volume = {61},
  issue = {1},
  pages = {906--919},
  numpages = {0},
  year = {2000},
  month = {Jan},
  publisher = {American Physical Society},
  doi = {10.1103/PhysRevE.61.906},
  url = {https://0-link-aps-org.pugwash.lib.warwick.ac.uk/doi/10.1103/PhysRevE.61.906}
}

@article{Freasier01011976,
author = {B.C. Freasier and D. Jolly and R.J. Bearman and},
title = {Hard dumbells: Monte Carlo pressures and virial coefficients},
journal = {Mol. Phys.},
volume = {31},
number = {1},
pages = {255--263},
year = {1976},
publisher = {Taylor \& Francis},
doi = {10.1080/00268977600100201},
URL = {https://doi.org/10.1080/00268977600100201},
}

@article{singer_monte_1990,
	title = {Monte {Carlo} study of fluid–plastic crystal coexistence in hard dumbbells},
	volume = {93},
	issn = {0021-9606},
	url = {https://doi.org/10.1063/1.459139},
	doi = {10.1063/1.459139},
	number = {2},
	urldate = {2025-04-23},
	journal = {J. Chem. Phys.},
	author = {Singer, Sherwin J. and Mumaugh, Ruth},
	month = jul,
	year = {1990},
	pages = {1278--1286},
}

@article{allen_molecular_1987,
	title = {A molecular dynamics study of the hard dumb-bell system},
	volume = {60},
	issn = {0026-8976},
	url = {https://doi.org/10.1080/00268978700100301},
	doi = {10.1080/00268978700100301},
	number = {2},
	urldate = {2025-04-29},
	journal = {Molecular Physics},
	author = {Allen, M.P. and Imbierski, A.A.},
	month = feb,
	year = {1987},
	pages = {453--473},
}

@article{kowalik_free_2008,
	title = {The free energy of hard dimer solids revisited},
	volume = {354},
	issn = {0022-3093},
	url = {https://www.sciencedirect.com/science/article/pii/S0022309308004262},
	doi = {10.1016/j.jnoncrysol.2008.06.050},
	number = {35},
	urldate = {2025-04-23},
	journal = {J. Non-Cryst. Solids},
	author = {Kowalik, M. and Wojciechowski, K. W.},
	month = oct,
	year = {2008},
	pages = {4354--4358},
}

@article{sweatman_self-referential_2009,
	title = {The self-referential method for linear rigid bodies: {Application} to hard and {Lennard}-{Jones} dumbbells},
	volume = {130},
	shorttitle = {The self-referential method for linear rigid bodies},
	number = {2},
    pages = {024101},
	journal = {J. Chem. Phys.},
    url = {https://doi.org/10.1063/1.3039190},
	author = {Sweatman, M.B. and Atamas, A. and Leyssale, J.-M.},
	year = {2009},
}

@article{wojciechowski_nonperiodic_1991,
	title = {Nonperiodic solid phase in a two-dimensional hard-dimer system},
	volume = {66},
	url = {https://0.link.aps.org/doi/10.1103/PhysRevLett.66.3168},
	doi = {10.1103/PhysRevLett.66.3168},
	number = {24},
	urldate = {2025-04-25},
	journal = {Phys. Rev. Lett.},
	author = {Wojciechowski, K. W. and Frenkel, D. and Brańka, A. C.},
	month = jun,
	year = {1991},
	pages = {3168--3171},
}

@article{nagle_new_1966,
	title = {New {Series}-{Expansion} {Method} for the {Dimer} {Problem}},
	volume = {152},
	url = {https://link.aps.org/doi/10.1103/PhysRev.152.190},
	doi = {10.1103/PhysRev.152.190},
	number = {1},
	urldate = {2025-06-19},
	journal = {Phys. Rev.},
	author = {Nagle, John F.},
	month = dec,
	year = {1966},
	pages = {190--197},
}

@article{Frenkel_cell,
    author  = "D. Frenkel",
    title = "Simulations: The dark side",
    year    = "2013",
    journal = "Eur. Phys. J. Plus",
    volume  = "128",
    url = "https://link.springer.com/article/10.1140/epjp/i2013-13010-8",
    pages   = "10"
}

\onecolumngrid
\pagebreak
\beginsupplement

\section*{Supplementary Information}

\subsection{Expanding the CPx family of close packed hard sphere dumbbells structures}
\label{sec:CPx}

In this section we expand upon the set of close-packed hard sphere dumbbell structures proposed by \citeauthor{vega_solidfluid_1992} \cite{vega_solidfluid_1992}. We demonstrate the rich variety of structures which can be constructed (and which are sampled by our NS simulations) within this family.

We start by defining a base layer of dumbbells arranged in a close packed configuration. The lower sphere in each dumbbell is arranged into a hexagonal 2D lattice in the $x$-$y$ plane. One can consider a rectangular unit cell in this layer with unit cell lengths  $a_1= \sigma$ and $a_2= 2\sigma\sqrt{3}$. Spheres within this unit cell are positioned at fractional coordinates of $(0,0)$ and $(0.5,0.5)$. Here we work with an unspecified diameter $\sigma$ of the constituent hard spheres for generality, noting that we set $\sigma=1$ for all simulations reported in the main text.

\begin{figure}[h]
\includegraphics[scale=0.7]{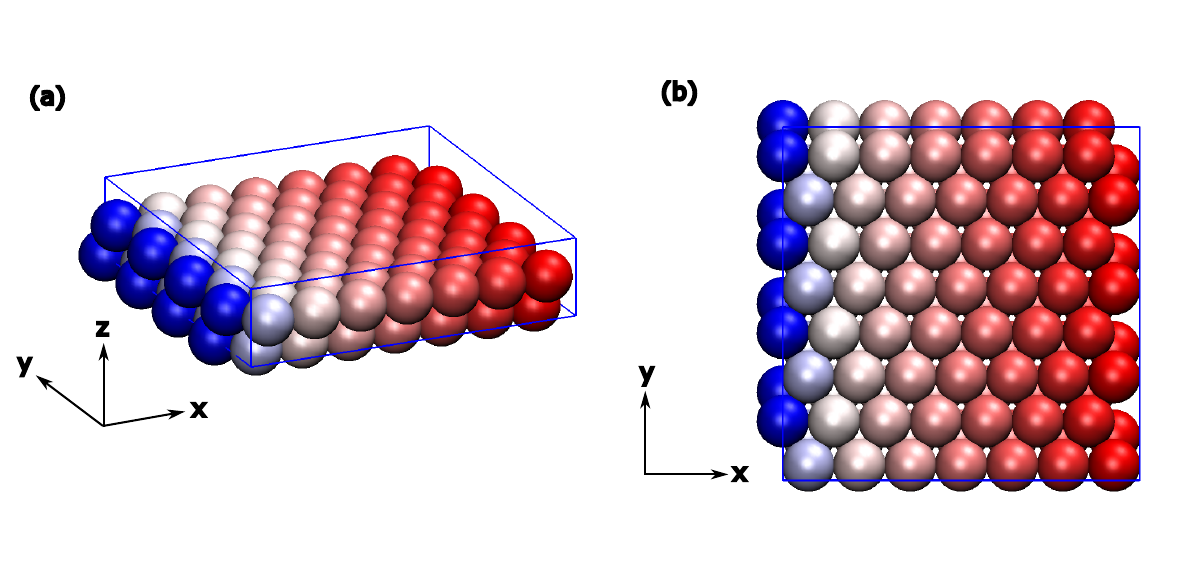}	
\caption{A close packed bilayer of hard sphere dumbbells with bond length $L$ equal to the constituent sphere diameter $\sigma$. A perspective view indicating the choice of coordinate axis is shown in (a). A top-down projection is shown in (b). Spheres are coloured according to their $x$-coordinate. Adjacent spheres of the same hue belong to the same dumbbell. }
\label{fig:SI_one_bilayer}
\end{figure}

To achieve maximal packing the \emph{bond vector} connecting to the upper sphere in each dumbbell must be tilted at an angle of $\theta = \arcsin{\left(L/\sigma\sqrt{3}\right)}$ to the vertical. Restricting ourselves to bond vectors which lie in the $y$-$z$ plane this produces a $\emph{bilayer}$ of two spheres layers, illustrated for $7\times 4$ rectangular unit cells and $L=\sigma$ in figure \ref{fig:SI_one_bilayer}. 

The two possible bond vectors that exist within the $y$-$z$ plane are as follows.

\begin{eqnarray*}
	\vec{b}_r &=& \left(0, -L \sin{\theta}, L \cos{\theta}\right), \\
	\vec{b}_l &=& \left(0, +L \sin{\theta}, L \cos{\theta}\right). 
\end{eqnarray*}
	
Here the (lower case) subscripts $r$ and $l$ indicate if the dimer tilts toward the left or right when viewed along the positive $x$ direction.

3D structures can be built by stacking further bilayers on top of the first, translated in the $y$-$z$ plane. We define the \emph{layer shift vectors} which connect the upper spheres of a layer to the lower spheres of the layer above. In order to achieve maximal packing the lower sphere in each layer must be centred above one of the hollow sites formed by the upper spheres of the previous layer. By symmetry there are only two non-equivalent choices for the position of this layer relative to the upper spheres of the previous layer, as is familiar from stacking of hard spheres. These correspond to the layer shift vectors

\begin{eqnarray*}
	\vec{b}_R &=& \left(0, -\sigma \sqrt{3}, \sigma\sqrt{2/3}\right), \\
	\vec{b}_L &=& \left(0, +\sigma \sqrt{3}, \sigma\sqrt{2/3}\right), 
\end{eqnarray*}
where the (upper case) subscripts now indicate a right or left tilt of the layer shift vector.

\begin{figure}[t]
\includegraphics[scale=0.7]{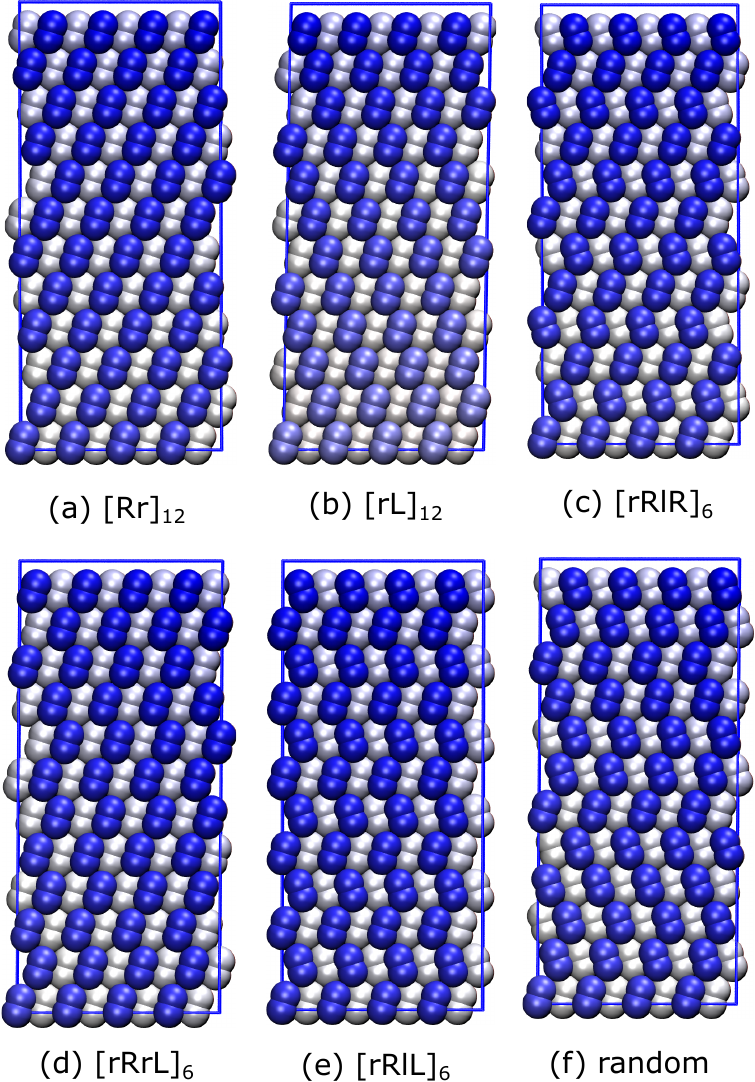}	
\caption{Representations of close packed dumbbell structures for $L=\sigma/2$. Unit cells with 12 bilayers are shown, viewed along the $-x$ axis with respect to the geometry of figure \ref{fig:SI_one_bilayer}. Structures (a) to (c) are the CP1, CP2 and CP3 structures described by \citeauthor{vega_solidfluid_1992}~\cite{vega_solidfluid_1992}. Structures (d) and (e) are two structures of equal complexity to CP3 which might be named CP4 and CP5. Structure (f) is a randomly generated sequence of bond vector and layer shifts equivalent to [rRlRlLlLlRrRlRlLrLrRlRlR]$_1$. Dimers are coloured according to their $x$ coordinate.}
\label{fig:SI_basic_CPs}
\end{figure}

We use the subscripts of the bond and layer shift vectors to denote which is selected for each layer. For example $[rR]_n$ (or equivalently $[lL]_n$) denotes $n$ bilayers in which the projections of both the dumbbell bond vector and layer shift vectors onto the $y$-axis are of the same sign. This is the CP1 dumbbell structure of \citeauthor{vega_solidfluid_1992}, and reduces to $[ABC]_n$ (fcc) packing of the constituent hard spheres for $L=\sigma$. The CP2 structure would be described as $[rL]_n$, i.e. all dumbbells are aligned and with layer shift vectors tilted in the opposite direction. This reduces to $[AB]_n$ (hcp) packing of the constituent spheres when $L=\sigma$. The CP3 structure of \cite{vega_solidfluid_1992} corresponds to $[rRlR]_n$ (or $[lLrL]_n$). This structure has a $z$ periodicity of 6 bilayers, corresponding to the 12 layer sequence $[ABCBCABABCAC]_n$ of constituent spheres when $L=\sigma$. Here the bond directions alternate between bilayers but the layer shift vector is unchanging. Representations of the structures CP[1-3] for $L=\sigma/2$ are shown in figure \ref{fig:SI_basic_CPs}(a) to \ref{fig:SI_basic_CPs}(c).  

To our knowledge no other sequences have been considered in the literature when studying the phase diagram of hard sphere dumbbells. Two more short periodicity sequences that might be considered are $[rRrL]_n$ in which the bond vectors are always in the same direction but the layer shift vector alternates, and the sequence $[rRlL]_n$ in which both bond vectors and layer shift vectors alternate between layers. These structures (which might be named CP4 and CP5) are also visualised in figure \ref{fig:SI_basic_CPs}(d) and \ref{fig:SI_basic_CPs}(e) respectively.

Many other structures with larger periodicity are of course possible and may be entropically favourable, analogous to stacking disordered close packing of hard spheres. An example of a randomly generated sequence of dimer and layer shifts with a periodicity of 12 bilayers is shown in figure \ref{fig:SI_basic_CPs}(f).

\begin{figure}[t]
\includegraphics[scale=0.7]{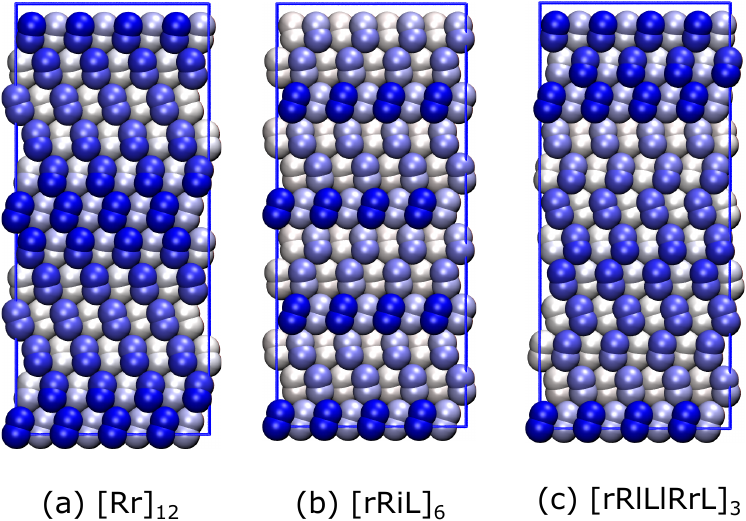}	
\caption{Representations of close packed dumbbell structures for $L=\sigma/2$. Unit cells with 12 bilayers are shown, viewed along the $-x$ axis with respect to the geometry of figure \ref{fig:SI_one_bilayer}. In each case the plane which contains the bond vectors rotates by $\pi/3$ around the $z$ axis between every bilayer. Dimers are coloured according to their $x$ coordinate.}
\label{fig:SI_helical_CPs}
\end{figure}

As noted in the main text, the restriction of the bond vectors to lie in the $y$-$z$ plane is not necessary. Structures with identical packing fraction can be built with bond vectors that are rotated around the $z$-axis at multiples of $60^{\circ} $ from the $y$-$z$ plane, such that the bond vector lies above the bridge site between two hollow sites in the plane formed by the lower spheres of a dumbbell layer. See figure \ref{fig:Pnma_crystal}(d) of the main text for an illustration. This choice of bond vector rotation introduces an additional degree of freedom to how layers are stacked and allows for a helical progression of the bond vector from layer to layer. Three examples of such structures are shown in figure \ref{fig:SI_helical_CPs}. 

Examples of all structures types described above emerge in our nested sampling calculations, however, with the number of dumbbells limited to either 16 or 32 we expect that structures with longer stacking periodicities were not captured.

\subsection{A second family of close packed hard sphere dumbbell structures}
\label{sec:CPx-interdig}

As described in the main text section \ref{sec:midbonds} our NS calculations also sampled close packed configurations which cannot be described as any specific sequence of bond vectors, layer shifts or bond vector rotations within the stacking model described above.

Instead these structures are built of layers in which the in-layer dimensions of the rectangular unit cell are $a_1 = \sigma$ and $a_2 = \sigma \sqrt{\alpha}/3 $, where 
\[
\alpha = 6-\lstar^2+2 \sqrt{2} \lstar\Delta, 
\]
and $\lstar=L/\sigma$ is the dimensionless bond length and 
\[
\Delta = \sqrt{3- \lstar^2}.
\]
 
Note that in contrast to the structures in the previous section, here the dimension of the 2D unit cell depends on $L$. 

There are again two dumbbells per unit cell of the constituent layers, with the lower sphere of one dumbbell occupying the corner of the unit cell. The lower sphere of the second dumbbell is offset from the lower sphere of the first by

\[
\vec{r}_o = \left(\sigma/2, \sigma \beta, \sigma \gamma / \alpha^{3/2} \right)
\]

where 

\[
\beta = \frac{(2-\lstar^2)}{2}\sqrt{\frac{3}{\alpha}},
\]

and

\[
\gamma = \sqrt{3} \left[
 \sqrt{2}\left(6+5\lstar^2-2\lstar^4\right) +
 \left(10 \lstar - \lstar^3\right)\Delta
 \right].
\]

\begin{figure}[t]
\includegraphics[scale=0.7]{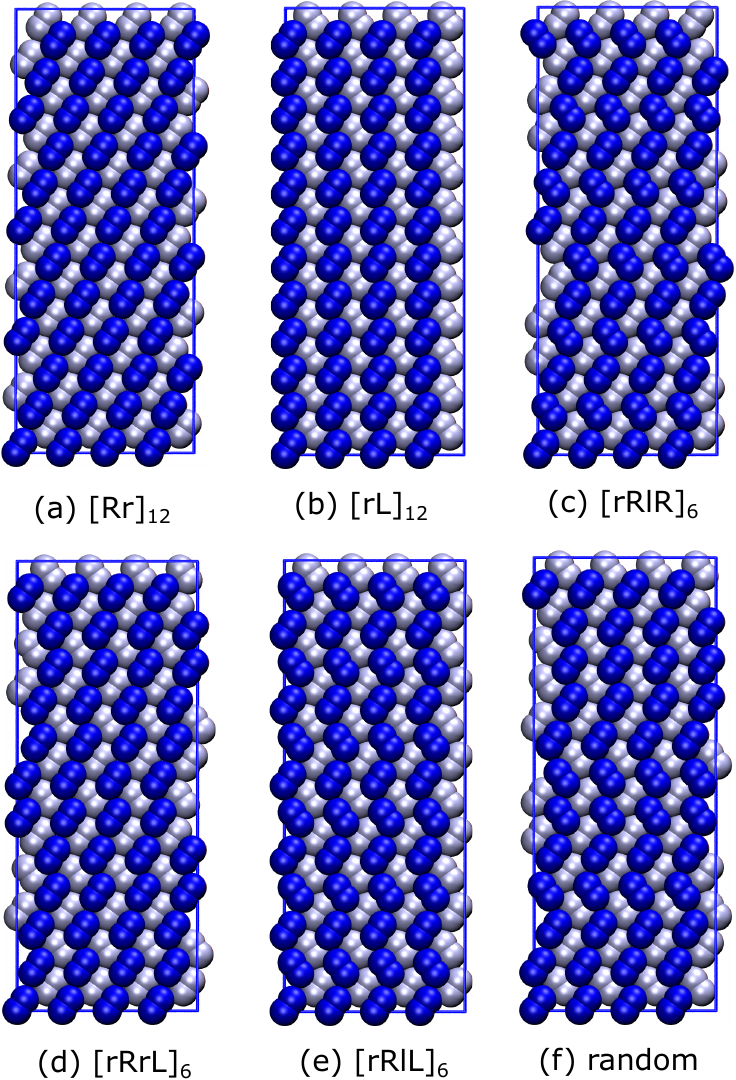}	
\caption{Representations of interdigitated close packed dumbbell structures for $L=\sigma/2$. 12 bilayers are shown, viewed along the $-x$ axis with respect to the geometry of figure \ref{fig:SI_one_bilayer}. Note that only (b) and (e) have periodicity in the stacking direction commensurate with the number of layers shown. Structures (a) to (c) are analogous to CP1, CP2 and CP3 and might be labelled as CP1i, CP2i and CP3i. Similarly structures (d) and (e) might be labelled as CP4i and CP5i. Structure (f) is a randomly generated sequence of bond vector and layer shifts equivalent to [rLrRrLlRrRlRlLrRrLrLrRrL]$_1$. Dimers are coloured according to their $x$ coordinate.}
\label{fig:SI_interdig}
\end{figure}

These layers can then be stacked into 3D structures in an analogous manner to the CPx structures, noting that the $z$-component of $\vec{r}_{0}$ results in the two dimers in each layer being interdigitated with each other. 

There are two possible bond vectors for dimers within a layer which are:

\begin{eqnarray*}
	\vec{b}_r &=& \frac{L}{\sqrt{3\alpha}}\left(0, 
	- \lstar - \sqrt{2}\Delta,
	\sqrt{2} \lstar + 2\Delta
	\right), \\
	\vec{b}_l &=& \frac{L}{\sqrt{3\alpha}}\left(0, 
	+ \lstar + \sqrt{2}\Delta,
	\sqrt{2} \lstar + 2\Delta
	\right).
\end{eqnarray*}
Similarly the two choices of layer shift (vector from the upper two spheres in a layer to the lower two spheres in the next) are:

\begin{eqnarray*}
	\vec{b}_R &=&  \frac{\sigma}{\sqrt{3\alpha}}\left(0, 
	- \lstar - \lstar \sqrt{2}\Delta, 
	\left[\sqrt{2}\left(18 -3 \lstar^2 - \lstar^4\right) + \left(18+\lstar - 5\lstar^3\right)\Delta\right]/\alpha
	\right), \\
	\vec{b}_L &=& \frac{\sigma}{\sqrt{3\alpha}}\left(0, 
	+ \lstar + \lstar \sqrt{2}\Delta, 
	\left[\sqrt{2}\left(18 -3 \lstar^2 - \lstar^4\right) + \left(18+\lstar - 5\lstar^3\right)\Delta\right]/\alpha
	\right). 
\end{eqnarray*}

Example structures corresponding to simple stacking sequences are shown in figure \ref{fig:SI_interdig}. The structure illustrated in figure \ref{fig:Pnma_crystal} of the main text corresponds to figure \ref{fig:SI_interdig}(e). We name these structures by analogy to the equivalent non-interdigitated structure, adding a "i" suffix to indicate that the layers are interdigitated. For example, figure \ref{fig:SI_interdig}(a)-(e) represent structures we label as CP1i to to CP5i. 

Note that when $L=\sigma$ these structures consist of a close packed structure in which the constituent spheres of the dumbbells form 2D hexagonal layers, but with a different bond arrangement to the equivalent $L=\sigma$ CPx structures. In this limit they constitute a further realisation of the aperiodic phase. However, away from $L=\sigma$ the structures are distinct.

It should also be noted that only stacking sequences with an equal number of right and left shifts will result in structures that can be represented by periodic unit cells with a small number of layers. Examples include panels (b) and (e) in figure \ref{fig:SI_interdig}. Other structures will have long or no periodicity in the stacking direction. For example, the structure shown in  \ref{fig:SI_interdig}(a) will have a stacking periodicity of $n$ layers only if 
\[
\frac{n\sqrt{3}}{2}  \left(\lstar  + \lstar\sqrt{2}\Delta \right)
\]
is an integer. In this case the same structure can easily be represented by a monoclinic unit cell of two dumbbells and is therefore accessible to our nested sampling. However, long/random stacking sequences such as those shown in figure \ref{fig:SI_interdig} would likely be inaccessible.

\begin{table}
\begin{tabular}{|c|ccccc|cc|}
\hline
$P\sigma^3/k_{B}T$ & CP1 & CP2 & CP3 & CP4 & CP5 & CP2i & CP5i \\
\hline
\hline
        20 &  0.6386(7) &  0.6387(8) &   0.638(1) &  0.6382(9) &  0.6389(6) &  0.6410(4) &  0.6405(4)\\  
        25 &  0.6643(4) &  0.6650(8) &  0.6653(4) &  0.6647(8) &  0.6645(5) &  0.6668(8) &  0.6662(5)\\  
        30 &  0.6831(7) &  0.6835(8) &  0.6834(7) &  0.6833(8) &  0.6831(7) &  0.6843(6) &  0.6837(5)\\  
        35 &  0.6969(6) &  0.6970(8) &  0.6968(9) &  0.6970(5) &   0.697(1) &  0.6980(3) &  0.6975(4)\\  
        40 &  0.7076(6) &  0.7091(3) &  0.7073(5) &  0.7072(6) &  0.7079(5) &  0.7084(3) &  0.7076(4)\\  
        45 &  0.7157(4) &  0.7158(5) &  0.7162(4) &  0.7156(4) &  0.7155(7) &  0.7163(4) &  0.7159(5)\\  
        50 &  0.7226(4) &  0.7227(4) &  0.7232(2) &  0.7226(5) &  0.7225(7) &  0.7229(4) &  0.7231(4)\\  
        55 &  0.7280(2) &  0.7279(4) &  0.7284(5) &  0.7282(3) &  0.7277(3) &  0.7284(8) &  0.7288(4)\\  
        60 &  0.7329(7) &  0.7329(2) &  0.7328(4) &  0.7331(7) &  0.7326(5) &  0.7332(5) &  0.7330(3)\\  
        65 &  0.7371(3) &  0.7371(6) &  0.7370(5) &  0.7371(6) &  0.7368(3) &  0.7372(3) &  0.7369(6)\\  
        70 &  0.7401(2) &  0.7403(2) &  0.7403(3) &  0.7403(3) &  0.7406(2) &  0.7404(4) &  0.7405(2)\\  
\hline
\end{tabular}
\caption{Packing fraction as a function of reduced pressure for a sample of close packed structures with $L^{\star}=L/\sigma=0.5$. Uncertainties in the last digit are shown in parenthesis and represent a 95\% confidence interval.}
\label{tab:cpfpl0.5}
\end{table}

\subsection{Packing fraction as a function of pressure for close packed structures at $L=0.5$}
\label{sec:CPpf}

Table \ref{tab:cpfpl0.5} compares the packing fraction of the polytypes CP1-CP5, and the two interdigitated polytypes (CP2i and CP5i) with short periodicity in the stacking direction for a bond length of $L/\sigma = 0.5$. To within the uncertainty of our calculations, the structures CP1-CP5 have identical packing fractions at all pressures studied. At pressures $P\sigma^3/k_{B}T > 50$ the two interdigitated structures also share this packing fraction. However, at lower pressures these two structures have a slightly (but statistically significant) higher packing fraction compared to the others. We attribute this difference to the dynamical nature of the non-interdigitated structures at low pressure. These can be observed to form stacking faults (relative to their initial layer sequence) during the simulations, disrupting their ideal structure. In contrast, the interlocked layers of the CP2i and CP5i structures do not. 

More detailed calculations will be required to determine whether the slightly higher density of these new structures leads to higher stability and if this trend is consistent throughout the range of $L/\sigma$ where these structures are stable with respect to the plastic crystal and aperiodic phases. We defer a more comprehensive investigation of this possibility to future work.

\end{document}